\documentclass[traditabstract]{aa}
\pdfoutput=1
\usepackage{color}
\usepackage[table]{xcolor}% http://ctan.org/pkg/xcolor
\usepackage{graphicx,subfigure}
\usepackage{amsmath}
\usepackage{txfonts}
\usepackage{footnote}
\makesavenoteenv{environmentname}
\usepackage{url}
\usepackage{float}
\usepackage{cite} 
\usepackage{natbib}
\usepackage{tikz}
\usepackage{ifthen}
\definecolor{orange}{RGB}{255,127,0}
\begin{document}

  \title{Searching for transits in the WTS with difference imaging light curves}

   \author{J. Zendejas
        \inst{1}$^,$\inst{2} 
        \and 
        J. Koppenhoefer\inst{2}$^,$\inst{1} 
        \and 
        R.P. Saglia\inst{2}$^,$\inst{1} 
        \and
        J.L. Birkby\inst{3}
        \and
        S.T. Hodgkin\inst{4}
        \and
        G. Kov\'{a}cs\inst{4}
        \and
        D.J. Pinfield\inst{5}
        \and
        B. Sip\H{o}cz\inst{5}
        \and
        D. Barrado\inst{6}$^,$\inst{7}
        \and
        R. Bender\inst{2}$^,$\inst{1}
        \and
        C. del Burgo\inst{8}
        \and
        M. Cappetta\inst{2}
        \and
        E.L. Mart\'{i}n\inst{9}
        \and
        S.V. Nefs\inst{3}
        \and
        A. Riffeser\inst{1}
       \and
        P. Steele\inst{2}
         }
      
      \institute{University Observatory Munich, Scheinerstrasse 1, 81679 M\"{u}nchen, Germany%\\
        \email{chicho@usm.uni-muenchen.de}\label{inst1}
        \and
        Max Planck Institute for Extraterrestrial Physics, Giessenbachstrasse, 85748, Germany\label{inst2}%\\
        \and
        Leiden Observatory, Leiden University, Postbus 9513, 2300 RA Leiden, the Netherlands\label{inst3}%\\
        \and
        Institute of Astronomy, Cambridge University, Madingley Road, Cambridge CB3 0HA,UK\label{inst4}%\\       
        \and 
        University of Hertfordshire, Centre for Astrophysics Research, Science and Technology Research Institute, 
        College Lane, AL10 9AB, Hatfield, UK\label{inst5}%\\
        \and
        Depto. Astrof\'{\i}sica, Centro de Astrobiolog\'{\i}a (INTA-CSIC), ESAC campus, PO Box 78, E-28691 Villanueva de la Ca\~{n}ada, Spain\label{6}%\\
        \and
        Calar Alto Observatory, Centro Astron\'omico Hispano Alem\'an, C/ Jes\'us Durb\'an Rem\'on, E-04004 Almer\'{\i}a, Spain\label{7}%\\
        \and
        Instituto Nacional de Astrof\'{i}sica, \'{O}ptica y Electr\'{o}nica, Luis Enrique Erro 1, Sta. Ma. Tonantzintla, Puebla, Mexico\label{8}%\\
       \and
        Centro de Astrobiolog\'ia (CSIC-INTA), Ctra. Ajalvir km. 4, 28850 Torrej\'on de Ardoz, Madrid, Spain\label{9}%\\
      }
      
\date {Manuscript accepted for publication in Astronomy \& Astrophysics: 02/10/2013}

\abstract{The Wide Field Camera Transit Survey is a pioneer program
  aimed to search for extra-solar planets in the near-infrared. The
  images from the survey are processed by a data reduction pipeline,
  which uses aperture photometry to construct the light curves. We
  produce an alternative set of light curves using the difference
  imaging method for the most complete field in the survey and carry
  out a quantitative comparison between the photometric precision
  achieved with both methods. The results show that difference
  photometry light curves present an important improvement for stars
  with J\,$>$\,16. We report an implementation on the box-fitting
  transit detection algorithm, which performs a trapezoid-fit to the
  folded light curve, providing more accurate results than the box-fit
  model.

We describe and optimize a set of selection criteria to search for
transit candidates, including a parameter calculated by our detection
algorithm, the $V$-shape parameter. The optimized selection criteria
are applied to the aperture photometry and difference imaging light
curves, selecting automatically the best 200 transit candidates from a
sample of $\sim$475\,000 sources. We carry out a detailed analysis in
the 18 best detections and classify them as transiting planet and
eclipsing binary candidates. We present one planet candidate orbiting
a late G-type star. No planet candidate around M-stars has been found,
confirming the null detection hypothesis and upper limits on the
occurrence rate of short period giant planets around M-dwarfs
presented in a prior study. We extend the search for transiting
planets to stars with J\,$\leq$\,18, which enabled us to set a more
strict upper limit of 1.1\,\%. Furthermore, we present the detection
of five faint extremely-short period eclipsing binaries and
three M-dwarf/M-dwarf binary candidates. The detections
demonstrate the benefits of using the difference imaging light curves
especially when going to fainter magnitudes.}


\titlerunning{Searching for transits in the WTS}
\authorrunning{J. Zendejas et al.}
\maketitle

\section{Introduction}
\label{sec:intro}

In recent years, the search for exo-planets has become an interesting
and exciting field in Astronomy. About thousand exo-planets have been
found since \citet{1995Natur.378..355M} detected the first planet
orbiting its host Main Sequence star. Measuring the host star radial
velocity variations represents one of the most successful techniques
to detect exo-planets, nevertheless only few parameters of the
planetary system can be determined with this method. This changes if
we search for a planet transiting its host companion. A transit occurs
when a planet blocks part of the surface from the star causing a
slight and periodic variation in its brightness, which can be detected
by a photometric analysis. This analysis provides information of the
planet and its host star and together with radial velocity
measurements, important physical parameters of the transiting planet
can be deduced, such as the mass and the radius.The first planetary
transit signal was reported in 2000
\citep{2000ApJ...529L..45C,2000ApJ...529L..41H} and since this
discovery, a significant number (more than 300) of exo-planets have
been detected transiting their host star.

Recently, several transit missions and surveys have been designed to
find and characterize new exo-planets. The most exciting and
successful projects designed to detect periodic transits are the space
missions Kepler \citep{2010Sci...327..977B} and
CoRoT\footnote{Convection, Rotation and Planetary Transits}
\citep{2008ASPC..384..270A,2008IAUS..249....3B}. Kepler was launched
on March 6, 2009 to observe more than 150\,000 stars and it is
expected to find a large number of Earth-size planets and Super
Earths. On the other hand CoRoT was originally designed to find
exo-planets with properties comparable to rocky planets in our Solar
System. Nevertheless, in June 2013, it was announced the culmination
of the CoROT mission, after six years of successful operation.

Earth-like planets are particularly interesting because if they
revolve in the habitable zone of their host star
\citep{1993Icar..101..108K}, the environment may be adequate to
support liquid water on the surface of the planet, which is believed
to be a key for the development of life. Cool and low-mass M-dwarf
stars are the most promising candidates to find Earth-like planets and
Super-Earths in the habitable zone. Due to their low effective
temperature ($T_{eff}$), the habitable zone of these stars is located closer to
them, therefore the change in their brightness caused by a planet
orbiting within this region is more evident. For instance, an
Earth-size planet orbiting a 0.08\,M$_{\odot}$ star produces a transit
of 1\,\% depth \citep{2009ApJ...698..519K}, a similar effect occurs
when a Jupiter radius planet blocks a Sun-like star.

Searching for transiting planets at near-infrared (NIR) wavelengths
provides important advantages to detect transiting planets around
M-dwarfs, since the peak of the Spectral Energy Distribution (SED) of
these stars falls in this spectral range. Several projects are
dedicated to study transiting planets around M-dwarf, such as APACHE
\citep{2012MNRAS.424.3101G}, PTF/Mdwarfs \citep{2012ApJ...757..133L}
and TRAPPIST \citep{2011Msngr.145....2J}. However, so far there are
only two transit projects focused on finding exo-planets around cool
and low-mass stars at NIR wavelengths, the MEarth project and the Wide
Field Camera
\footnote{Wide Field Camera, see
  \url{http://www.jach.hawaii.edu/UKIRT/instruments/wfcam/}} (WFCAM)
Transit Survey (WTS).  The MEarth project
\citep{2009AIPC.1094..445I,2012AJ....144..145B} is a transit survey
that operates since 2008 with 8 independent 0.4\,m robotic telescopes
located at the Fred Lawrence Whipple Observatory on Mount Hopkins,
Arizona, and is soon expected to include eight additional telescopes
in the Southern hemisphere. The survey monitors individually $\sim$
2000 nearby ($<$33\,pc) M-dwarfs in the NIR and is designed to detect
exo-planets as small as 2\,R$_{\oplus}$. On the other hand, the WTS is
a pioneer project operated since 2007 with observations from the
United Kingdom InfraRed Telescope (UKIRT) that stands out for its
particular aims and methodology. A brief description of the WTS is
summarized in Section \ref{sec:wts}.

Traditionally, aperture photometry (AP) has been the standard
technique to produce light curves in transit surveys. In 1996, a new
method to study crowded fields by optimal image subtraction was
presented by \citet{1996AJ....112.2872T} and subsequently improved by
\citet{1998ApJ...503..325A}. This method (usually called "difference
imaging"-DI) was initially developed to study microlensing events in
crowded fields. However, since the majority of transit survey targets
are crowded fields (e.g. Galactic plane), image subtraction photometry
has become an important tool to search for planetary transits
\citep{2010A&A...509A...4P}. In the past, some authors have carried
out comparisons between different photometric techniques. For
instance, \citet{2007A&A...470.1137M} used the data from a ten-night
observing campaign from 4 different ground-based telescopes to develop
a quantitative test by comparing the photometric precision of 3 different
photometry algorithms: AP, PSF-fitting photometry and image
subtraction photometry. They compare the photometric precision as a
function of the apparent visual magnitude for all photometric
techniques. Due to the several factors involved in the observations
(which influence directly in the measurement), such as size of the
telescope, instruments or atmospheric conditions, the quality of the
light curves clearly varies depending on the location of the
observations. For all cases presented in \citet{2007A&A...470.1137M},
the best Root Mean Square (RMS) was achieved by image subtraction
photometry, in some cases a difference of up to 4 mmag is observed for
bright objects. On the other hand, AP and PSF fitting photometry show
significant variations of the photometric precision achieved by each
telescope. This discrepancy suggests that the precision obtained by a
certain photometric technique may depend on the characteristics of the
survey, i.e. a particular method might produce different results
depending on the observing conditions. In this work, we carry out a
similar analysis by comparing the photometric precision of the WTS
light curves obtained by DI and AP. 

Large sets of light curves usually show systematic effects that can be
associated with the atmospheric extinction, detector efficiency or
simply PSF changes on the detector. The $sysrem$ algorithm proposed by
\citet{2005MNRAS.356.1466T} has been widely tested and it is commonly
used in transit surveys
\citep{2007A&A...476.1357S,2006MNRAS.373..231P} to decrease the number
of systematics in light curves. To reduce these effects
in our sample, we apply the $sysrem$ algorithm and subsequently
include the results in the comparison analysis.

Due to the large number of light curves in
transit surveys, an efficient detection algorithm is needed to speed
up the identification of planet candidates. Shortly, after the
discovery of the first planet transiting its host star, several
algorithms have been developed. For instance,
\citet{2001A&A...365..330D} presented an algorithm that uses a
multi-frequency Fourier fit to model periodic astronomical time
series. \citet{2002A&A...391..369K} presented a box-fitting algorithm
based on least squares fit of step functions (BLS) to analyze stellar
photometric time series. This algorithm has shown significantly better
results than previous works and it has become a popular tool to search
for exo-planets in transit surveys. Recently the Transit Planet Search
(TPS) algorithm \citep{2010SPIE.7740E..10J} has been developed to be
part of the Kepler Science Processing Pipeline to search for transit
planets, which is able to achieve super-resolution detection
statistics.

False-positives and false-detections are common problems that make
difficult the search for exo-planets in transit surveys. A
false-detection can be caused if the light curve holds a significant
number of systematic outliers, which can produce fake signals, whereas
a false-positive is associated to real variability of the flux from
the host star, (e.g. eclipsing binary systems or intrinsic
  variable stars). Although false-positives and false-detections have
conceptually different origins, for practical reason, in this work
both scenarios are referred as false-positives. Nowadays, large scale
transit surveys require strategies to efficiently weed out
false-positives in candidate samples and reduce the number of light
curves inspected by visual examination. Several authors have suggested
methods to reduce the number of false-positives and facilitate the
selection of the best candidates. For instance,
\citet{2006AJ....132..210B} proposed a series of selection criteria
based on a $\chi^2$-minimization equivalent to the analytic solutions
provided by BLS method. Later on, \citet{2009ApJ...695..336H}
presented selection criteria divided in different steps, which include
the signal-to-pink noise ratio \citep{2006MNRAS.373..231P}, the number
of data points in the light curves, a magnitude limit and exclusion of
sources with alias periods between 0.99 and 1.02 days or less than 0.4
days. Nevertheless, the majority of selection criteria only remove
false-positives not related to real astrophysical variability. In this
study we propose a selection criterion, which has the ability of
excluding false-positives taking into account elements from the
transit detection algorithm, as well as a new criterion named
"$V$-shape parameter" that is designed to recognize automatically
eclipsing binary systems.

The structure of this paper follows the next outline: In Section
\ref{sec:data_analysis} we describe the WTS and summarize the image
reduction pipeline. In this section we also give a description of the
DI analysis and describe the procedure of the light curve
extraction. A quantitative comparison between the photometric
precision of light curves obtained by AP and DI techniques is
presented in Section \ref{sec:comparison}. The Section
\ref{sec:transit_detection} is dedicated to present our transit
detection algorithm and the $V$-shape parameter obtained from the
implementations made on the BLS algorithm. Sections
\ref{sec:selection_criteria} and \ref{sec:opt_selec_crit} present our
selection criteria and the optimization of the parameters used to
detect planet candidates on the WTS light curves. In Sections
\ref{sec:candidatesFGK} and \ref{sec:candidatesM} we show the
candidates that pass our selection criteria and a detailed physical
characterization of the candidates. We present in Section
\ref{sec:applications} other applications of the WTS DI light curves,
such as the detection of ultra-short period and detached M-dwarf
eclipsing binaries. Finally, we summarize our results in Section
\ref{sec:conclusion}.

\section{Data analysis \& methodology}
\label{sec:data_analysis}
\subsection{The Wide Field Camera Transit Survey}
\label{sec:wts}
Low-mass main-sequence stars of spectral type M are the most abundant
stars in the Galaxy, representing about 75$\%$ of the total stellar
population \citep{2007AsBio...7...85S}. In addition, M-dwarfs present
certain properties that make them ideal targets to search for rocky
planets \citep{2007AsBio...7...30T}. Motivated by this, the WTS was
initially developed to monitor and search for transiting planets for
the first time in the NIR. Since the transit technique is based on
relative photometry, the survey can be performed in poor weather
conditions, hence WTS is conducted as a back-up project, operating
when the observing condition are not suitable (seeing $>$ 1 arcsec)
for the main program of the UKIRT Infrared Deep Sky Survey
(UKIDSS). The survey was originally assigned about 200 nights at the
3.8m UKIRT equipped with the WFCAM, which consists of 4 Rockwell
Hawaii-II arrays with 2048$\times$2048 pixels in each panel that cover
a field of view of 0.75 square degrees with a resolution of 0.4
arcsec/pixels. The 4 detectors are distributed geometrically at the
corners of a square with an auto-guider located at the center of the
frame. This array is usually called $pawprint$, a complete observation
sequence of the WTS consists of 8 $pawprints$ (a-h) and each one is
built up from a nine-point jitter pattern of 10s. An entire field is
completed in about 15 min, i.e, the WTS light curves have an average
cadence of four data points per hour. Since the dimension and
separation of the detectors have approximately the same size ($\sim$
13 arcmin), a uniform target field can be achieved by observing the 8
$pawprints$.  Four fields were selected seasonally to be observed (RA
= 03, 07, 17 and 19h) periodically during a year, thereby the WTS
guarantees a reasonable continuous observations
campaign. Nevertheless, this work is only dedicated to study the
RA\,=\,19h field, which has been observed until May 2011 with about
1145 epochs and contains $\sim$\,475\,000 sources, of which
$\sim$\,113\,000 have magnitudes J\,$\leq$\,18. All observations for
the WTS are done in the J-band ($\mathrm{\lambda_{eff} \approx}$
1200nm). For more details about the WTS we refer to
\citet{2013MNRAS.tmp.1446K}. The image reduction procedure will be
described in the next section.

\begin{figure*}[!ht]
     \begin{center}
       \subfigure[]{%
            \label{fig:first}
            \includegraphics[width=0.4\textwidth]{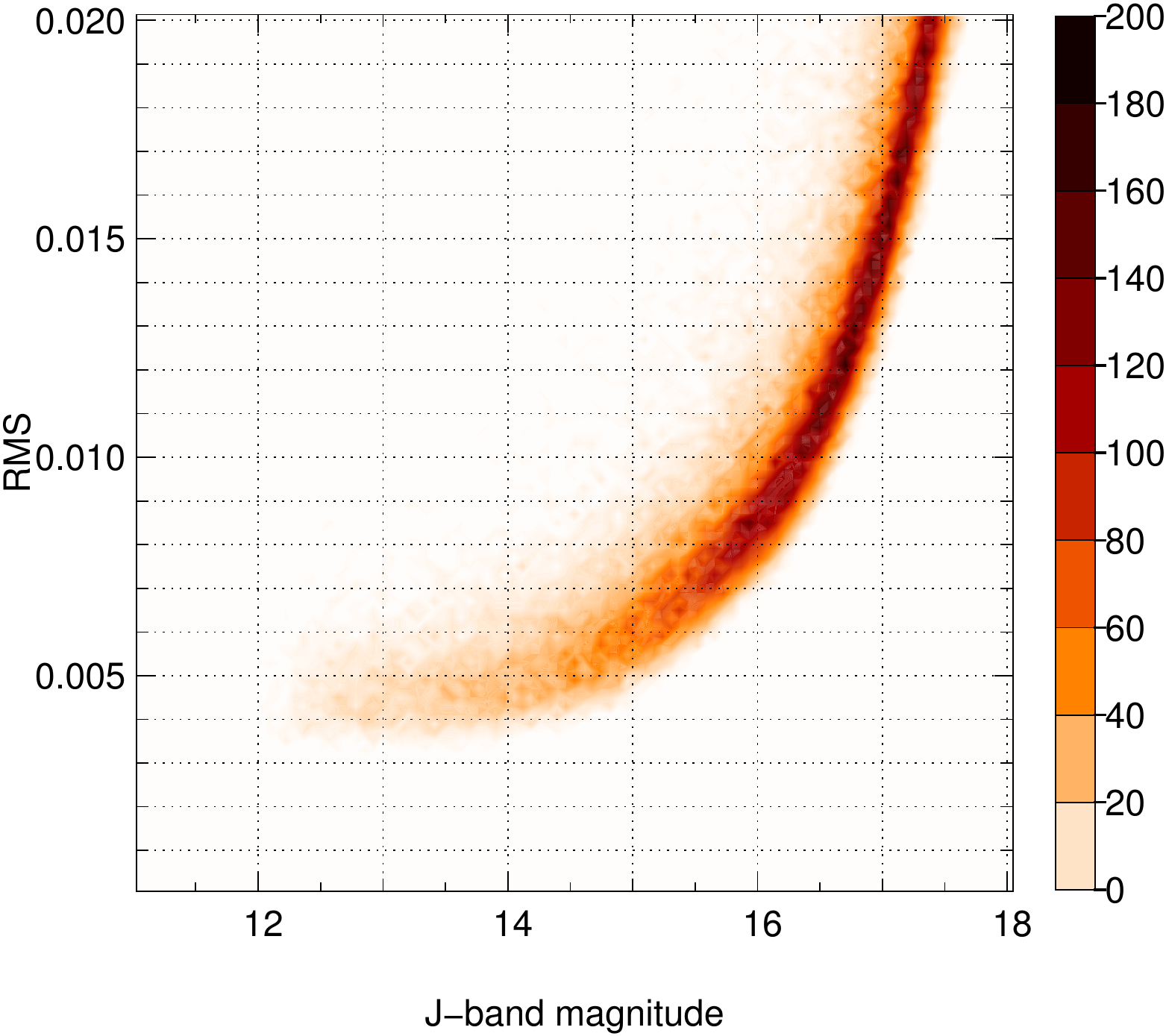}
        }%
        \subfigure[]{%
           \label{fig:second}
           \includegraphics[width=0.4\textwidth]{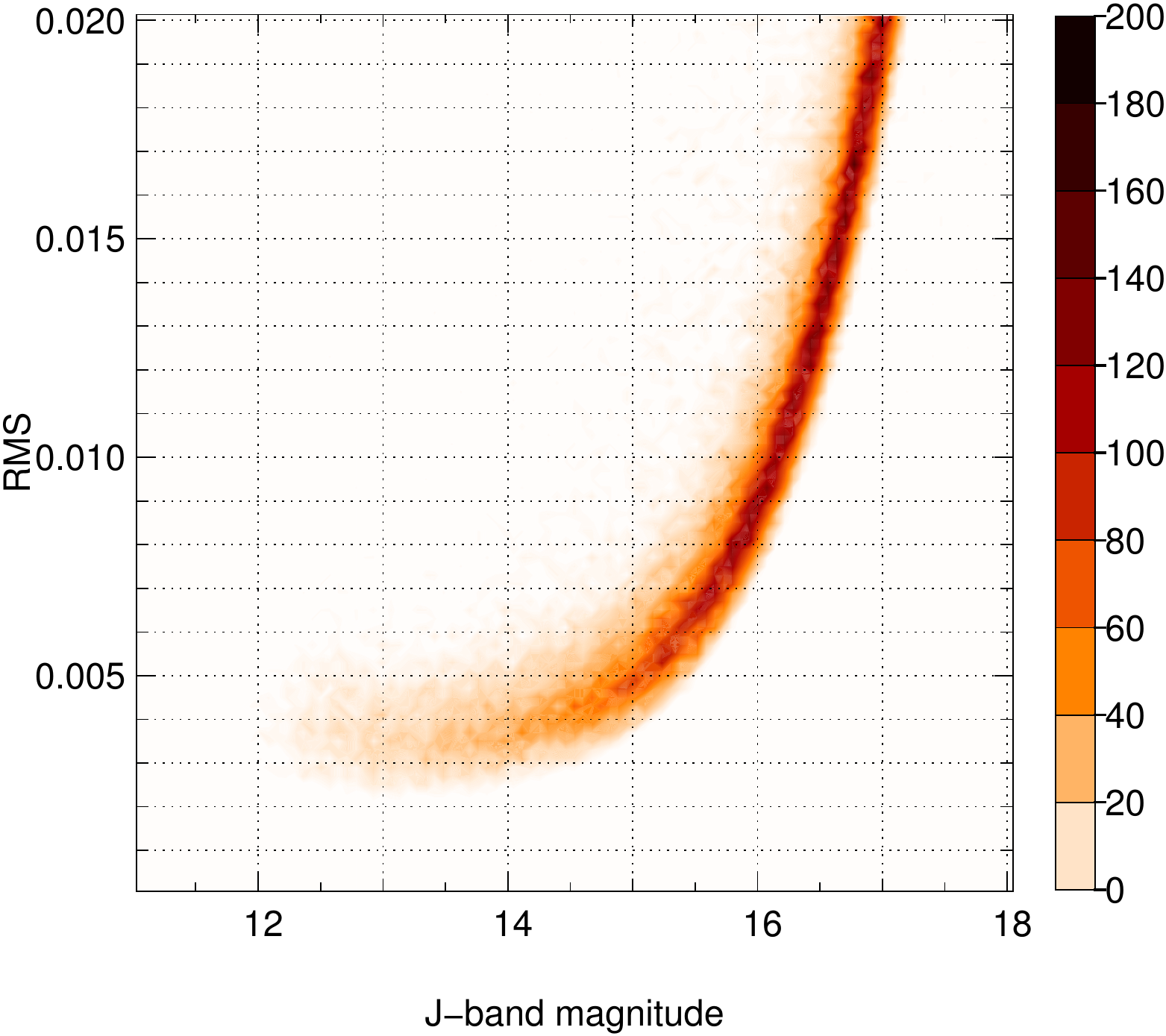}
        }

         \end{center}
    \caption{Quantitative comparison between different
        photometric analysis. The figure shows The RMS of the DI and
        AP light curves (panels a and b, respectively) as a function of
        the J-band magnitude. The RMS corresponds to the measurements
        obtained after removing systematic effects. The plot is
        displayed in density of data points in a scale of 100 bins.
     }%
   \label{fig:cam_usm_sys}
\end{figure*}

\subsection{Image reduction pipeline}
\label{sec:image_reduction}
Due to the large amount of data collected by the WTS, a pipeline to
process the images automatically is required. The J-band images from
the WTS are reduced by the image reduction pipeline from the Cambridge
Astronomical Survey
Unit\footnote{\url{http://casu.ast.cam.ac.uk/surveys-projets/wfcam}}
(CASU), which is used to process all images from the WFCAM. The image
reduction pipeline is based on the work developed by
\citet{1985MNRAS.214..575I} and later modified and adapted to the
Isaac Newton Telescope (INT) Wide Field Survey (WFS)
\citep{2001NewAR..45..105I} and subsequently to the Monitor project
\citep{2007MNRAS.375.1449I}. The pipeline includes the following
steps: De-biassing and trimming, non-linearity correction, bad pixel
replacement, flatfielding, defringing and sky subtraction. A thorough
description of all the steps can be found in
\citet{2001NewAR..45..105I}. Astrometry and photometry are calibrated
using bright stars in the field-of-view from the 2-Micron All-Sky
Survey (2MASS) \citep{1994Ap&SS.217...11K} catalog (see
\citealt{2009MNRAS.394..675H}). Particularly, the astrometric
calibration plays an important role in the DI technique, since a
precise alignment of data frames is crucial to success with this
method. The astrometry is described by six coefficient linear
transformations allowing for scale, rotation, shear and coordinate
offset corrections. The pipeline also provides master catalog and
light curves, which are constructed by the AP method, using a series
of soft-edge-apertures that account for the fractional area of a pixel
included in the aperture, with the addition of a simultaneous
redistribution of flux from nearby stars. More detailed description of
the light curves and catalog can be found in
\citet{2013MNRAS.tmp.1446K}. In the next section we describe the DI
method and the process of the light curves extraction.

\subsection{Difference Imaging Analysis}
\label{sec:diffima}
In addition to the standard WTS light curves (AP) generated by the
CASU pipeline, we alternatively produce a second set of light curves
by using DI photometry. According to \citet{1998ApJ...503..325A}, the
method operates on a reference image, which is the combination of the
best seeing images from the data set ($\sim$\,20 in our case). On
average the seeing range of the images used to construct the reference
frames is 1.18 to 1.39 arcsec. The reference frame is convolved with a
kernel to match the seeing of each single image, resulting in a
convolved reference image. A difference image is obtained by
subtracting the convolved reference image from each single image.

Finding the optimal kernel that matches the seeing of two frames with
different PSFs represents a crucial and complex problem during the DI
process. \citet{1998ApJ...503..325A} proposed a method, in which the
optimal kernel is approximated using a superposition of N kernel base
functions, which are constituted of 2-dimensional Gaussian functions
modulated with a polynomial of order $p_i$. The
expression for the optimal kernel is :\\

\begin{equation}
  K(u,v) = \sum_{i=1}^{N} {exp \left[-\frac{u^2 +
        v^2}{2\sigma_i^2}\right]} \sum_{j=0}^{p_i} \sum_{k=0}^{p_i-j}
  {a_{ijk}u^jv^k},
\end{equation}

where $u$ and $v$ are the pixel coordinates of the kernel bitmap,
which has the same pixel size as the images, $a_{ijk}$ are the
coefficients from the decomposition of the kernel using basis of
functions and $\sigma_i$ is the variance related to the Gaussian
distribution.  To calculate the kernel, we use four base functions
($N$=4) with $\sigma_i$= 1,2,3 and 0.1, while the degrees of the
associated polynomials $p_i$ are 6,4,2 and 0, respectively. The kernel
size is 11x11 pixel and we consider a 1st order background polynomial
to account for background difference. All free parameters, such as the
$a_{ijk}$ coefficients and the parameters associated to the background
polynomial are determined by minimization of the following
expression:\\

\begin{equation}
  \chi^2=\sum_{x,y}{\frac{1}{\sigma_{x,y}^2}}[\{(R(x,y) \otimes
    K(u,v)\} + B(x,y) -S(x,y)]^2,
\end{equation}

where $\sigma_{x,y}^2$ is the variance of a Gaussian distribution used
to approximate the images Poisson statistics, S(x,y) is a single
image, R(x,y) is the reference frame and B(x,y) is the polynomial
surface function that accounts for background differences.Variations
of PSF over the detector are a common problem in the DI technique. In
order to reduce this effect during the estimation of the kernel, we
divide the images in subfields and calculate the kernel in each
subfield. In our case we divide the images in 10x10 subfields with a
size of 200x200 pixels.\\

To achieve an optimized set of difference images, we tested several
parameters. For each set, we extract the light curves and measure the
photometric precision to verify the quality of the sample. During the
testing process, we found that the light curves precision is
significantly improved if we mask bright or faint stars while the
difference images are produced. Two sets of difference images are
created to guarantee the best quality of the light curves. In a first
set, we mask all sources with magnitude $\leq$ 16, which provides an
improvement for objects fainter than this threshold. The second set is
processed by masking faint objects, i.e. all sources that hold
magnitude $>$ 16, which results in an improvement for bright stars.

\subsection{Light curve extraction}
\label{sec:lc_extraction}
From the difference images, we are able to measure the differential
flux of each source. Adding the value measured in the reference image,
the total flux for each single star can be estimated. Although
differential fluxes are relative easy to measure in the difference
images, because all constant sources are removed, estimating the
fluxes in the reference frame is more difficult, especially for
objects that have close neighbors. We measure the flux in the
reference frame using iterative PSF-photometry. This technique is very
successful to measure flux accurately in crowded fields. The method
uses bright and isolated stars to extract the PSF. In a first step an
initial estimation of the flux of each star is measured from the
extracted PSF. In subsequent iterations, all nearby stars are removed
before measuring the flux of a particular source. This process
continues until all fluxes converge, using in each iteration the
improved flux measured in the previous step. The fluxes measured in
difference images are also estimated by PSF-photometry. The PSF is
obtained from the convolved reference image, using the same stars
employed to estimate the flux in the reference frame, which are a
representative sample of stars in each field. Although the fluxes in
the difference images certainly could be estimated by using a
different photometric technique (e.g. aperture photometry), since the
stellar crowd in the field is eliminated, we have chosen
PSF-photometry to measure the fluxes because this method is not
affected by dead pixels and does not require aperture corrections,
which might lead to a wrong evaluation of the flux. Finally, the light
curves are normalized to one and barycentrically time corrected using
the formula of \citet{1982QB51.3.E43M43..}. The process of extracting
the light curves is applied to both sets (one optimized for bright
sources and one optimized for faint sources, see previous section). We
obtain the optimized set by choosing the light curve with the better
photometric precision for each source.

\section{Quality of the difference imaging light curves and comparison
  with the aperture photometry method}
\label{sec:comparison}
In this section we compare the quality of the WTS light curves
produced by AP (from the CASU pipeline) and DI. A quantitative
comparison between the photometric precision of both sets of light
curves is performed by calculating the RMS of each
  single light curve from the two photometric analysis. During this
process we clip all 4$\sigma$-outliers, while clipping 3 and
5$\sigma$-outliers provides similar results. Note that this step is
only for the purpose of calculating the RMS and is not a final
operation on the light curves.\\

\subsection{Sysrem algorithm}
\label{sec:sysrem}
An algorithm to remove systematic effects in large sets of light
curves from photometric surveys was proposed by
\citet{2005MNRAS.356.1466T}. The algorithm, called $sysrem$, has shown
the capability of improving considerably the photometric precision of
the data set by removing systematics related to the detector
efficiency, PSF variations over the detector or effects associated
with the atmospheric extinction
\citep{2009A&A...506..431M,2007MNRAS.375.1449I}. The algorithm
searches for systematics that consistently appear in many sources of
the sample, hence $sysrem$ has the ability to remove effects without
any prior knowledge of the origin of the effect.\\

In order to improve the quality of the light curves, we consistently
apply the $sysrem$ algorithm to DI and AP light curves. Note that
\citet{2007MNRAS.375.1449I} showed that the $sysrem$ algorithm does
not improve the precision of AP light curves by much and it might
additionally produce false variability from the residuals. In our case
we find a significant reduction of the scatter of constant light
curves for both DI and AP light curves. Any possible false variability
created by $sysrem$ will not lead to the detection of false-positive
candidates because of our conservative criteria applied in the
candidate selection process (see below).\\

The results are shown in Figure~\ref{fig:cam_usm_sys}, which represent
the RMS achieved by the AP and DI light curves (panel a and b
respectively) after the correction as a function of the WFCAM J-band
magnitude. The DI light curves reach a precision of 3.5 mmag for
bright objects in the range of 12 $<$ J $<$ 14, while the RMS of AP
light curves corrected by $sysrem$ algorithm reaches a precision 
$\sim$\,2.5 mmag in the same J-band magnitude interval. The plots show
that DI produces better results for faint objects (J $>$ 16), however
in the bright magnitude range, the quality of AP light curves is
slightly better.  For magnitudes larger than J\,=\,16, the DI light
curves show a much higher photometric precision than the AP light
curves. The RMS shows presents a difference up to 5\,mmag at
J\,=\,17\,mag and 15-20\,mmag at J\,=\,18\,mmag.\\

\begin{figure}[ht!]
  \centering
    \includegraphics[width=0.4\textwidth]{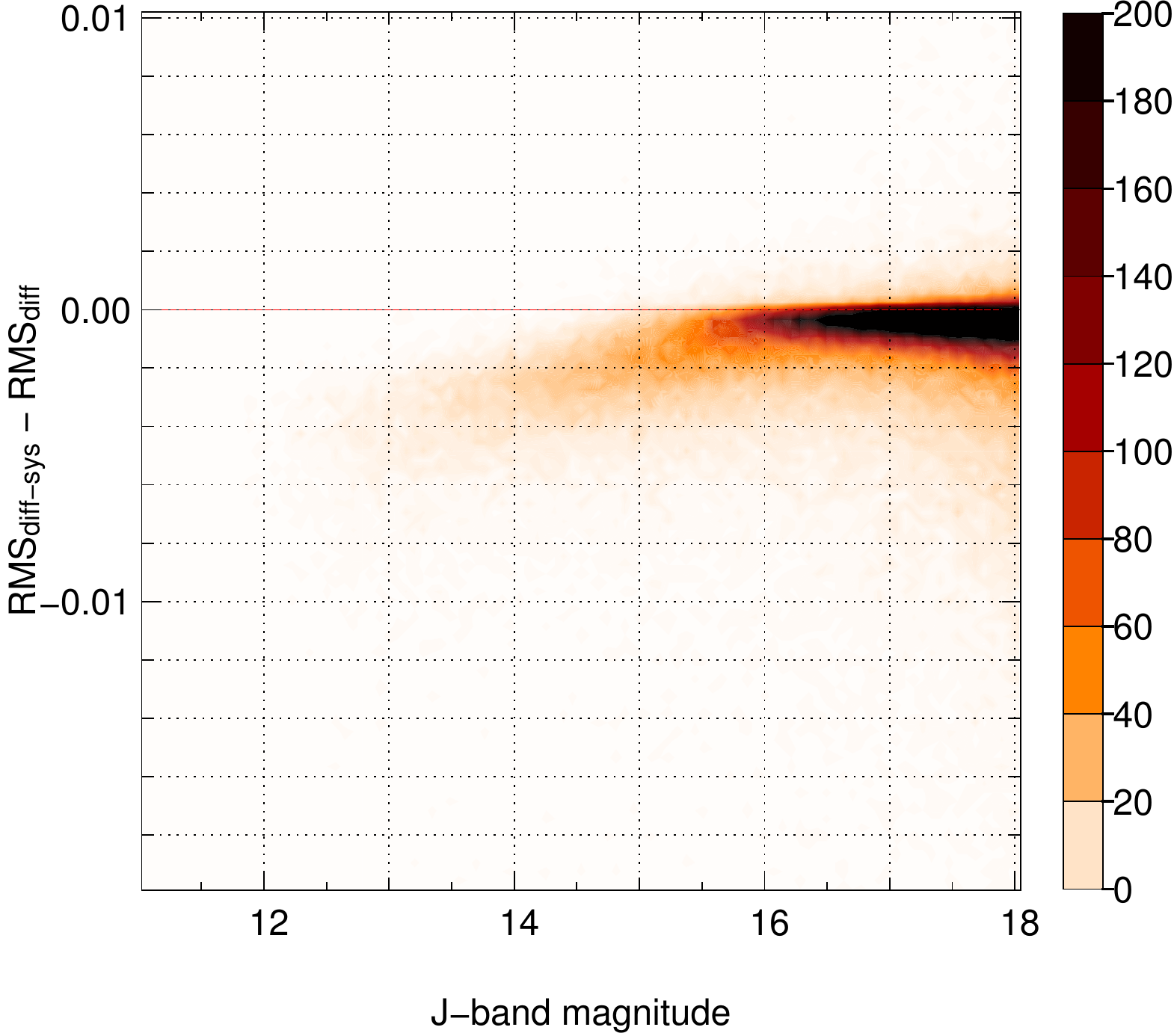}
    \caption{RMS difference between the DI light
        curves before and after applying $sysrem$ algorithm. The plot
        shows the improvement achieved in the photometric precision once
        systematic effects are corrected. The plot shows the density of
        data points distributed in 100 bins.}
  \label{fig:usm_sys_usm}
\end{figure}

\begin{figure}[h!]
  \centering
    \includegraphics[width=0.3\textwidth]{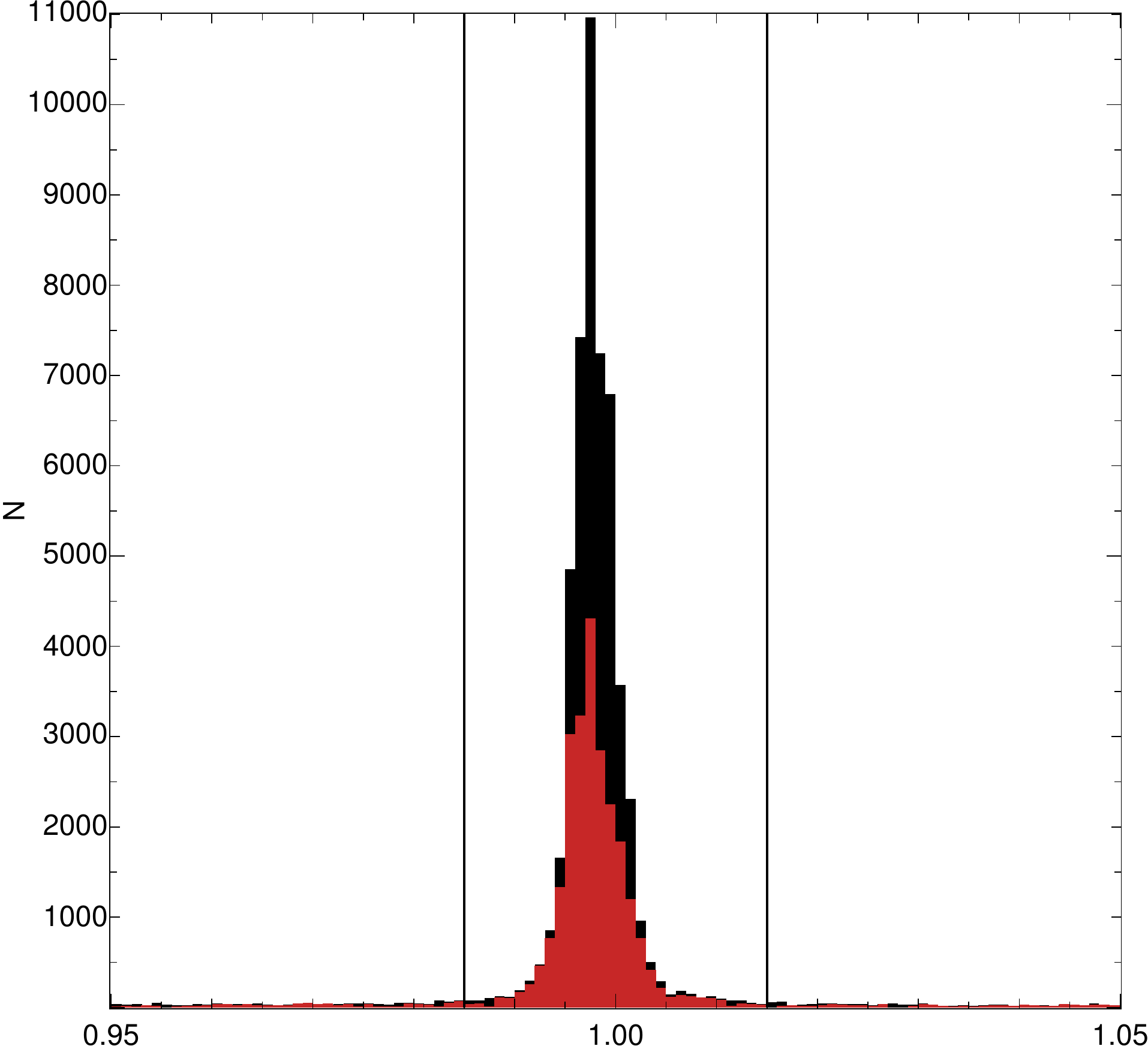}
    \caption{Histogram of periods found by our transit
        detection algorithm, before (black) and after (red) applying
        the $sysrem$ algorithm. The numbers of false detections
        between 0.985 and 1.015 day periods is reduced by a factor of
        2.}
  \label{fig:histogram_criteria}
\end{figure}   

These results contrast to previous studies, which compare the
photometric precision achieved with both methods. For example,
\citet{2007A&A...470.1137M} show that DI photometry achieves an equal
or better photometric precision compared to aperture and PSF
photometry for all magnitudes. However, these studies were done at
optical wavelengths (V-band) and a direct comparison to a
NIR survey (like the WTS) is not possible, since the detector
characteristics are different. Imperfect treatment of non-linearity
effects at the bright end could be one possible source for the
additional systematic noise that we observe in our DI light curves. Another
problem might be the non-homogeneous background, which is visible in
the WTS images. We can rule out that the effect is caused by a low
astrometric accuracy. Any shifts between the reference frame and the
single images would produce dipole-shaped residuals in the difference
image, contrary to this effect, bright sources show very symmetric
noise residuals in our difference images.\\

In order to show the capability of the $sysrem$ algorithm to improve
the photometric precision, we perform a similar quantitative analysis
(see above) on the DI light curves by comparing the RMS of the light
curves before and after applying the $sysrem$
algorithm. Figure~\ref{fig:usm_sys_usm} shows the RMS difference
between both sets of light curve as a function of the J-band
magnitude.  The result of the comparative analysis indicates a
significant improvement in the photometric precision of bright and
faint sources when the $sysrem$ algorithm is employed. A similar
result is observed in the AP light curves. Note that although
  applying the $sysrem$ algorithm results in an improvement of the
  photometric precision of the light curves, the capability of the
  algorithm to remove systematics effects is
  limited. Figure\,\ref{fig:histogram_criteria} shows the number of
  detected periods around the one-day alias period before and after
  using the $sysrem$ algorithm. In an ideal case, the algorithm should
  account for these effects and eliminate the alias peak. In our
  case, the number around the alias period is reduced by a factor
  $\sim$\,2 after applying the $sysrem$ algorithm.

\begin{figure*}[!ht]
     \begin{center}
       \subfigure[]{%
            \label{fig:first_err_corr}
            \includegraphics[width=0.4\textwidth]{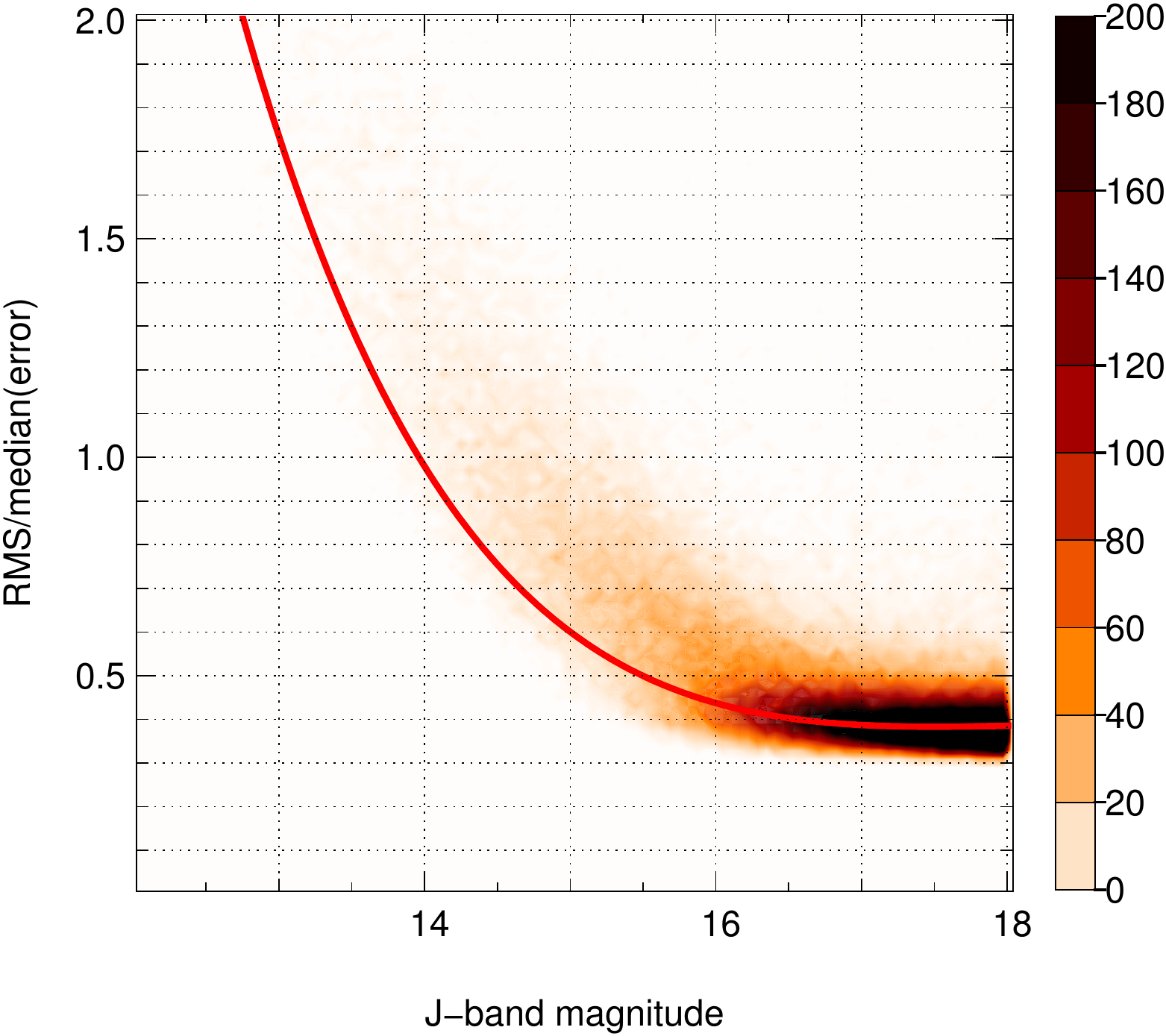}
        }%
        \subfigure[]{%
           \label{fig:second_err_corr}
           \includegraphics[width=0.4\textwidth]{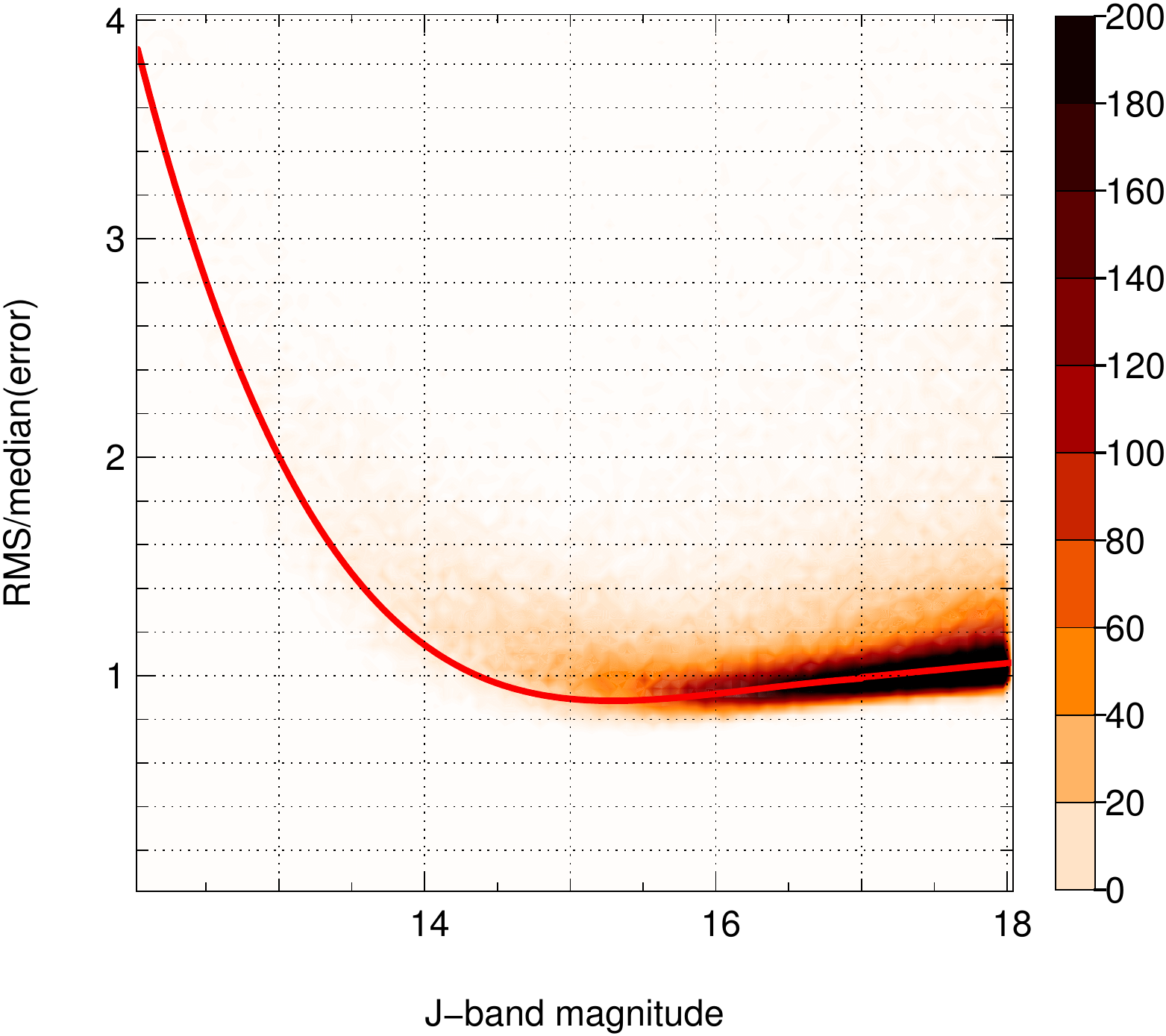}
        }

         \end{center}
    \caption{Data points distribution of the RMS divided by the median
      error from DI and AP light curves (panel a and b respectively) as a function of J-band
      magnitude. The solid-red line represents the polynomial used to
      scale the errors bars. The plot is displayed in
      density of data points in a scale of 100 bins.
     }%
   \label{fig:rms_median_ap_di}
\end{figure*}

\begin{figure*}[!ht]
     \begin{center}
       \subfigure[]{%
            \label{fig:first_norm_corr}
            \includegraphics[width=0.4\textwidth]{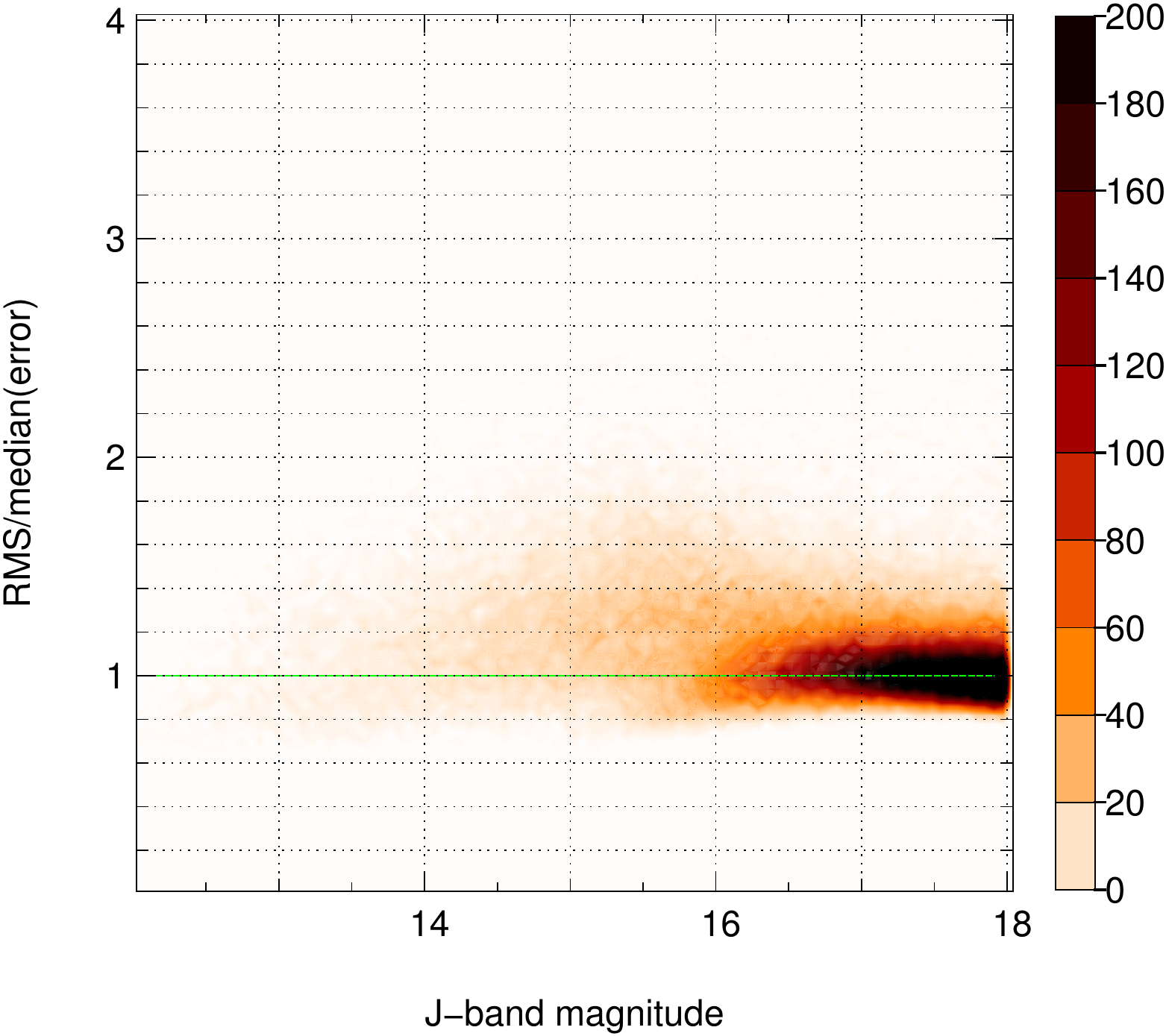}
        }%
        \subfigure[]{%
           \label{fig:second_norm_corr}
           \includegraphics[width=0.4\textwidth]{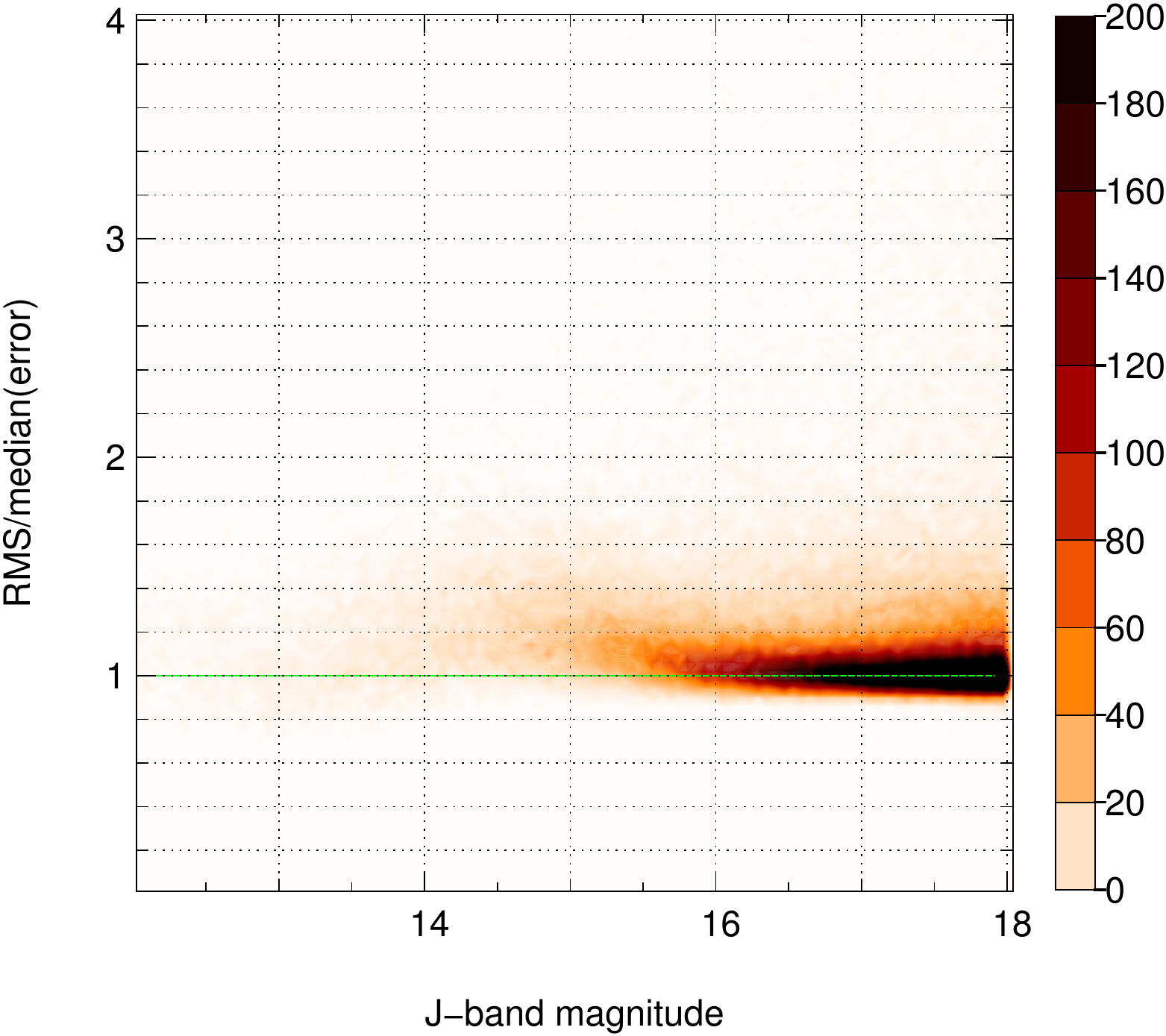}
        }

         \end{center}
    \caption{Distribution of data points RMS/median(error) after
      re-scaling the error bars on the (a)DI and (b)AP light curves.
     }%
   \label{fig:rms_median_norm_ap_di}
\end{figure*}

\subsection{Correction of the point-by-point errors derived from the individual images}
\label{sec:ebars}
After removing systematic effects, we carried out a routine visual
inspection over a random sample of light curves. We noticed that for
many light curves (typically for bright sources) the scatter of the
data points was much larger than the error bars. Usually error bars of
light curves are estimated by a pipeline taking into in account
different factors, such as the photon noise of the source, background
contribution and read-out noise. However, systematic effects caused by
PSF-variations or variation in noise level from the background are not
included. This seems to be the case of the WTS light curves, which
present a wrong estimation of the error bars, which is correlated to
the brightness of the object. A simple way to correct the size of the
error bars is to compare the RMS of the photometric measurements with
the error values, since the RMS is related to the scatter level and it
can be associated with the real error for non-variable objects. We
perform this test for DI and AP light curves dividing the RMS by the
median error calculated in each light curve. The results are shown in
Figure \ref{fig:rms_median_ap_di}, where this quotient is plotted as a
function of the J-band magnitude. If the error values were correct,
the RMS and median error should present similar values, therefore data
points in the plots should be distributed around 1. Nevertheless,
there is an evident discrepancy between both quantities, which is
reflected in the shape of the data point distributions. For our work,
it is important to correct the bad estimation on the error bars, since
some of our selection criteria (see below) and several parameters that
we estimate for our candidates later-on depend on the error bars. In
order to correct the error bars, we fit a polynomial (Figure
\ref{fig:rms_median_ap_di}) to the distribution of data points. The
polynomial provides a scale factor as a function of the magnitude,
which can be used to correct the whole sample. Note that unlike
replacing the errors obtained from the images by an estimation of the
scatter (RMS), scaling the error bars with a factor that is a function
of the brightness of the objects, we avoid to introduce an
overestimation of the errors, principally for variable sources, which
can present a significant scatter in the light curves. After scaling
the error bars, we perform the same test and show the results in
Figure \ref{fig:rms_median_norm_ap_di}, where we can see that the
distribution of data points clearly have been adjusted and they are
now located close to 1. Nevertheless, these figures present a second
higher RMS sequence for bright stars (14-16 magnitudes). We know from
the AP light curves (see \citealt{2013MNRAS.tmp.1446K}) that the WTS
data present a high level of red noise \citep{2006MNRAS.373..231P}
also correlated to the magnitude of the objects, being the bright
sources the most affected for this effect. Although the $sysrem$
algorithm is designed to filter out the red noise, there is a
component from the red/pink-noise that remains in the sample of light
curves, which can be observed in Figure \ref{fig:rms_median_ap_di},
where a significant scatter is visible in the distribution of the data
points. The fact that $sysrem$ cannot eliminate completely this
component of the red/pink noise may produce fake signals and
subsequently a large number of false-positives. Figure
\ref{fig:histogram_criteria} demonstrates that the remaining
systematics produce such effects, since a large number of objects fall
into the daily alias. Nevertheless, in Section
\ref{sec:selection_criteria} we introduce the selection criteria used
to detect planet candidate in the WTS -light curves, which have the
capability to provide a pure candidate sample, ruling out false
positive related to some of these systematics. On the other hand, in
this work we do not use the correlated noise to measure the transit
fitting significance. Therefore, the polynomial used to correct the
error bars does not take into in account the dispersion of data points
generated by the remaining red/pink-noise component.

\section{Light curves analysis and transit detections}
\label{sec:transit_detection}
\begin{figure}[h!]
  \centering
    \includegraphics[width=0.48\textwidth]{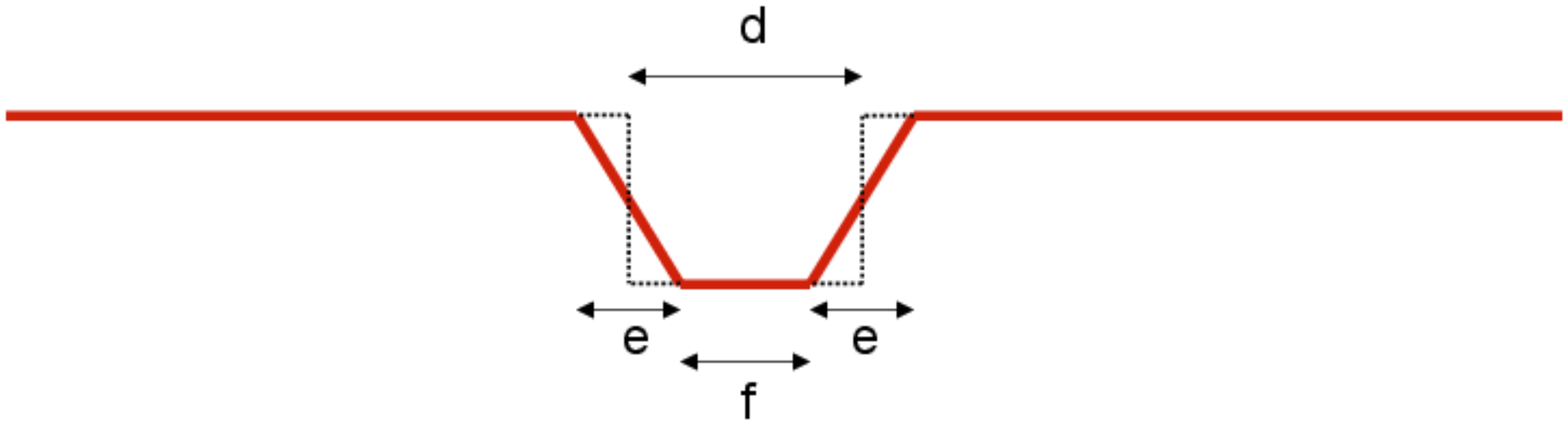}
  \caption{Geometry of the symmetrical trapezoid-fit.}
  \label{fig:v_shape}
\end{figure}

\begin{figure}[h!]
  \centering
    \includegraphics[width=0.4\textwidth]{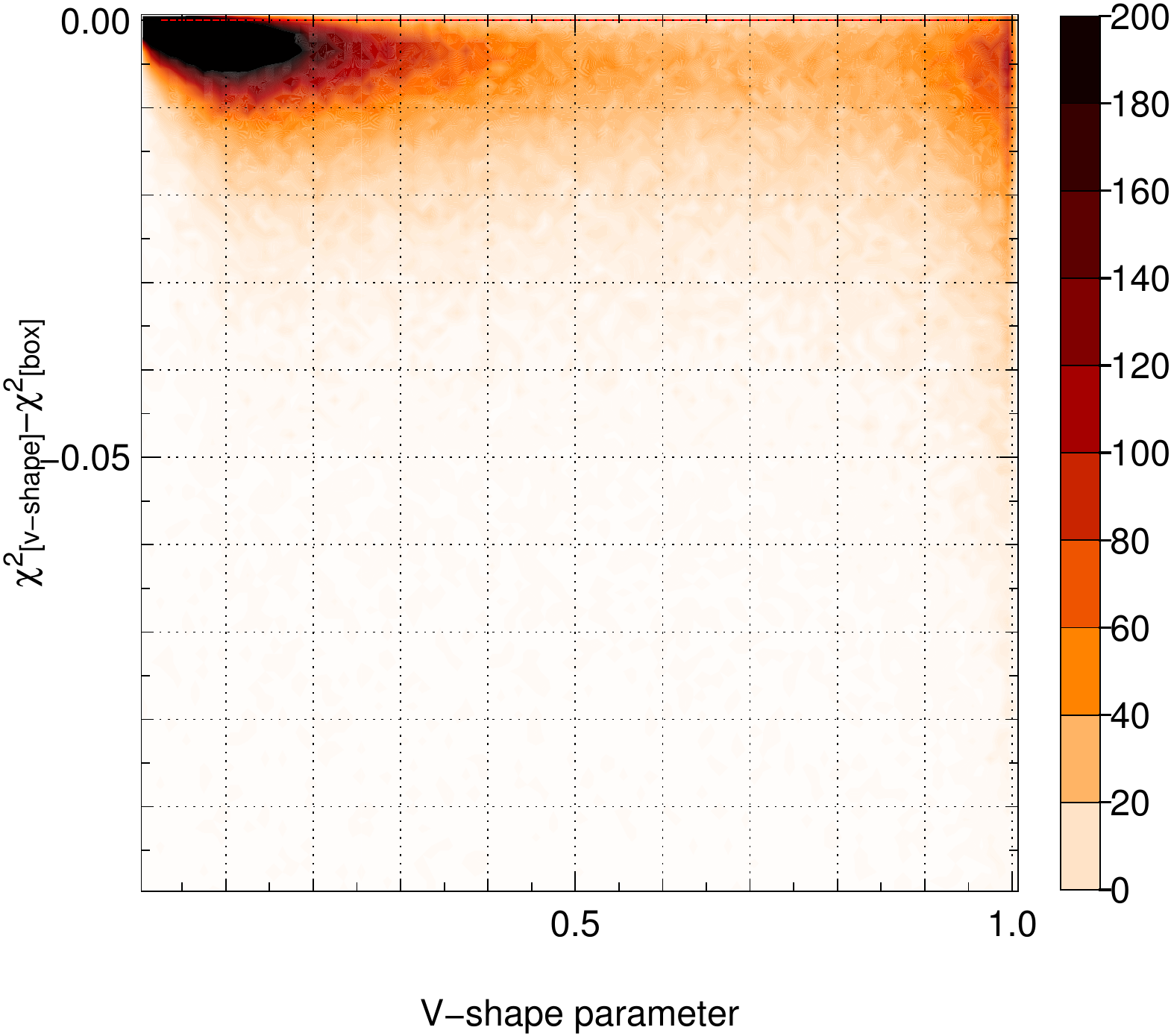}
    \caption{$\chi^2_{dof}$ comparison between trapezoid-fit and
      box-fit for transit detections. Since the
        box-fit is actually included in the trapezoid-fitting (i.e.,
        $V$=0), positives values are not expected in this plot. On the
        other hand, a significant improvement in the trapezoid-fit is
        achieved specially for higher values of $V$.}
  \label{fig:v_shape_chi}
\end{figure}

We detect transits in the WTS light curves using an algorithm that is
based on the BLS algorithm proposed by
\citet{2002A&A...391..369K}. Our modifications include a trapezoidal
re-fitting of the box-shaped eclipse found by BLS, where the
re-fitting is done by symmetrically varying the edges of the box while
keeping fixed the duration of the eclipse (``d''), which is measured
at half the transit depth (see Figure~\ref{fig:v_shape}). We emphasize
that the trapezoidal shape is only fitted once the standard parameters
provided by the box-fitting algorithm have been found (such as period,
transit duration and epoch), however, the eclipse depth may change. We
introduce the $V$-shape parameter:\\
\begin{equation}
  V =  \frac{2e}{f + 2e},
  \label{eq:v_equation}
\end{equation}        
where $e$ is the duration of the ingress/egress of the eclipse in
phase units and $f$ is the duration of the transit, i.e., the flat
part (see Figure \ref{fig:v_shape}). If $e$ is considerably smaller
than $f$, the $V$-shape parameter is close to 0 and the shape of the
fit is box-like. On the other hand, if $f$\,$\approx$\,0 and
$e$\,$\gg$\,$f$, the $V$-shape parameter is close to 1 and the eclipse
is ``$V$''-shaped. One of the advantages of our modification is that
the $V$-fitting results in a better estimate of the transit depth. In
addition, we use the $V$-shape parameter as a selection criterion to
reject grazing eclipsing binary systems which have generally very
large $V$ values (see next section).\\ We search for transit periods
in the range between 0.5 and 12 days using 100\,001 trial periods
equally distributed in 1/P. To speed-up the calculation time, the
folded light curves are re-sampled to 200 bins. The fractional transit
duration was tested between 0.006 and 0.1 phase units. For each input
light curve we detect the 5 best fitting periods with the BLS
algorithm and then perform the trapezoidal re-fitting for each of
them. We then select the period that has the lowest $\chi^2_{dof}$ of
the improved $V$-fit. Figure \ref{fig:v_shape_chi} shows the
difference between the reduced $\chi^2_{dof}$ of the trapezoid-fitting
and the box-fitting as a function of the $V$-shape parameter. The
trapezoid-fitting shows a significant improvement over the box-fitting
especially for high $V$ values.\\

\subsection{Selection Criteria}
\label{sec:selection_criteria}

Due to the large number of light curves in the WTS, it is necessary to
set up a number of selection criteria to automatize the selection of
candidates but efficiently reducing the number of false positives in
the survey. As an initial cut we removed all objects with magnitudes
J\,$>$\,18. Objects below J-band\,=\,16 are already difficult to
follow-up, nevertheless we decided to extend the magnitude cut
(J-band\,=\,17) used in \citet{2013MNRAS.tmp.1446K} in order to make
use of the improvement achieved by difference imaging light curves for
faint objects.\\ In addition, we reject objects for which the
detection algorithm found a period close to alias periods introduced
by the window function of the observing strategy. In particular we
exclude objects with periods in the ranges 0.485-0.515, 0.985-1.015,
1.985-2.015 and 2.985-3.015\,days. As an example,
Figure~\ref{fig:histogram_criteria} shows the high number of (false)
detections found around the one day alias period. For the sub-field
19g1 we additionally exclude a narrow period range 1.350-1.352\,days
due to a very high number of false positives in this range. Based on
our experience of previous works, we introduce six more selection
criteria:

\begin{enumerate}

 \item{{\textbf{\emph{S/N}:}} One of the most important criteria
     is the $S/N$ of the eclipse measured from the light curves. In the
     past, many authors have used different ways to calculate the
     $S/N$ and many different ways of utilizing it as a selection
     criterion. For instance \citet{2006AJ....132..210B} include the
     signal-to-white-noise in their selection criteria to set the
     threshold to $S/N$\,$\geq$\,10.  \citet{2009ApJ...695..336H}
     propose the same limit of $S/N$\,$\geq$\,10, but in their case
     the threshold corresponds to the pink noise
     \citep{2006MNRAS.373..231P}.  \citet{2013MNRAS.tmp.1446K} use the
     red noise to fix the detection limit, they suggest a
     signal-to-red noise of $S_{red}$\,$\geq$\,6. Our $S/N$ selection
     criterion accounts only for white noise.}

 \item{{\textbf{\emph{S/N-S/N$_{rem}$}}:} A large
     fraction of false positive detections are variable stars. To
     eliminate them, we use a new detection criterion, labeled
     $S/N-S/N_{rem}$, which is the difference of the $S/N$ found in
     the BLS analysis and the $S/N_{rem}$ found in a second pass of
     the algorithm after masking all points during the eclipse that
     has been detected in the first interaction. For a planet
     candidate $S/N-S/N_{rem}$ will be very high since the variability
     is confined to the transit phase. For variable stars,
     e.g. eclipsing binaries or sinusoidal variables, there is still
     variability left which will result in a low value of
     $S/N-S/N_{rem}$. Note that this criterion will eliminate the
     detection of systems with more than one transiting
     planet. However, since the WTS survey is only sensitive to
     periods smaller than ten days, we decided to only search for
     systems with a single transiting planet.}

\item{{\bf \# points:} Many light curves result in a
    high S/N detection but only very few points belong to the
    transit. We therefore require a minimum number of transit points
    in our candidate selection process. Due to the scheduling of the
    WTS, we do not require a minimum number of individual transits as
    an additional criterion, since even a small minimum number of
    transit points guarantees implicitly two transits or more.}

\item{{\textbf{\emph{V$_{shape}$}:}} One selection criterion that
    has not been used in previous studies is the $V$-shape parameter,
    which was defined in the previous section. The criterion acts as a
    filter to eliminate false-positives generated by eclipsing binary
    systems. An eclipsing binary would be characterized by a very
    V-shaped eclipse, i.e. with a high $V$ value.}

\item{{\bf depth:} Some detections show a very deep
    transit signal. A typical brightness dimming corresponding to a
    Sun-like star and a Jupiter-like planet is about 1\,\%. Transit
    signals that are much deeper are more likely to be eclipsing
    binary stars. Using a cut on the maximum allowed transit depth we
    reduce the number of false positive detections. Note that for
    Jupiter-sized planets around M-dwarfs, the transit depth can be
    even higher than 10\,\%. We therefore optimize the detection
    criteria for M-stars independently (see below).}
  
\item{{\bf transit duration:} In order to exclude
    candidates that show un-physically long eclipses we impose a limit
    on the fractional transit duration.}
  
\end{enumerate}

\subsection{Optimization of the selection criteria}
\label{sec:opt_selec_crit}

We optimize our selection criteria with Monte-Carlo simulations, where
we inject transit signals in the real light curves using stellar
parameter distributions (radius and mass) from the Besancon model of
the galaxy \citep{2003A&A...409..523R} and the limb-darkening
coefficients from \citet{2011A&A...529A..75C}. For each light curve of
the WTS we pick a random star from the Besancon model that has a
similar magnitude ($\Delta_{mag} \le$ 0.05) and draw a random period
in the range from 0.5 to 12\,days. Details about the transit injection
procedure can be found in \citet{2009A&A...494..707K}.\\ We split the
light curves in two data sets, one for F-, G- and K-stars and one for
M-stars. The optimization of the selection criteria was done
separately, since we expect some parameters to differ between both
data sets. For instance, the transit depth is generally larger for
planets orbiting M-dwarfs and the fractional transit duration is
smaller. In addition, the particular analysis of the M-dwarf sample
allowed us to derive an upper limit on the occurrence rate of
Jupiter-sized planets around low-mass stars (see Section
\ref{sec:candidatesM})\\ The M-dwarf selection is based on color cuts
in seven SDSS and WFCAM bands: g-r $\ge$ 1.6, r-i $\ge$ 0.9, i-z $\ge$
0.5, J-H $\ge$ 0.45 and H-K $\ge$ 0.17. These cuts have been derived
to include the majority of M-dwarfs selected by
\citep{2013MNRAS.tmp.1446K}. Based on these cuts we find 10\,375
M-stars brighter than 18\,mag in J-band. 4\,073 objects are brighter
than J=17 which is slightly less but still in reasonable agreement
with the number of objects selected in \citet{2013MNRAS.tmp.1446K} who
found 4\,600 M-dwarfs using an SED fitting approach. In the following
sections we report the optimized selection criteria and the detection
efficiency for both data sets and present and discuss the selected
candidates.\\\\

\section{Candidates detected around F-, G- and K-stars}
\label{sec:candidatesFGK}

For the simulated light curves we required the detected period to be
within 1\% of the simulated period, allowing also a value of half or
double the this period. On a computer cluster we ran in total 100
simulations. In each run we considered each light curve once for a
transit injection, resulting in about 1\,200\,000 simulated light
curves. In order to save computation time, we simulated only those
cases, in which the randomly drawn inclination vector results in a
visible transit signal. After running the simulations we optimized the
selection criteria presented above for the DI and AP light curves of
all F-, G- and K-stars. We allowed up to 100 detections on the
unmodified light curves on each data set. This number is strategically
selected, since it is small enough to allow a visual inspection of
each detected object, while being significantly larger than the
expected number of planet detections.\\ Tables \ref{tab:selec_diffima}
and \ref{tab:selec_aperture} list the optimized selection criteria for
the DI and AP light curves for F-, G- and K-stars and provide the
number of objects that remain after applying each of the selection
criteria. In this case, the fractional transit duration turned out to
be a useless criterion to detect candidates around these
stars. These selection criteria allow us to recover 10/26\% of
  the signals injected into the AP/DI light curves with S/N
  $\sim$\,11/18 (our minimum required S/N) and up to 80/80\% with S/N
  $\ge$\,30/40, respectively. The resulting total efficiencies are
  discussed in Section\,\ref{sec:candidates_disccusion}. Note that before
applying the magnitude limit, the number of light curves in the DI and
AP data sets differ at the 10\% level. This is because the object
detection in the DI analysis was going slightly deeper than in the AP
analysis.\\ In order to test whether the selection criteria differ
from one to another detector we initially optimized them for each of
the sub-fields independently but found almost identical values. We
therefore decided to use one single set of selection criteria for F-,
G- and K-stars in the the whole 19hrs field.\\

\begin{table}[h!]\bf
   \caption{Objects removed by the selection criteria from a original sample of 464873 DI light curves.}  
    \scalebox{0.85}{
     \begin{tabular}{l*{9}{c}}
      \hline 
      Criterion                     & Remaining objects & Removed objects & \%       \\\hline                        
      J\,$\leq$\,18                   &  102428         & 362445   & 76.26   \\  
      Removed alias period            &   72012         &  30416   & 29.69   \\  
      S/N\,$>$ \,18                   &    7080         &  64932   & 90.17   \\  
      S/N-S/N$_{rem}$\,$>$\,8         &    3391         &   3689   & 52.10   \\  
      Transit points\,$>$\,24         &     506         &   2285   & 85.08   \\  
      V$_{shape}$\,$<$\,0.6          &     288         &    218    & 43.08   \\  
      Depth\,$\leq$\,4\,\%            &     100         &    188   & 65.27   \\  
      Transit duration\,$\leq$\,0.5   &     100         &      0   & 0.00     \\
      \hline
    \end{tabular}
  }
  \label{tab:selec_diffima}
  \tablefoot{Number of DI candidates after applying all criteria (100).}
\end{table}

\begin{table}[h!]\bf
  \caption{Objects removed by the selection criteria from a original sample of 428928 AP light curves.}
   \scalebox{0.85}{
    \begin{tabular}{l*{9}{c}}
      \hline 
      Criterion                      & Remaining objects & Removed objects & \%     \\\hline                        
      J\,$\leq$\,18                   & 102428     & 326500  & 74.32 \\  
      Removed alias period            &  73201     &  29227  & 28.53 \\  
      S/N\,$>$ \,11                   &   5778     &  67423  & 92.11 \\  
      S/N-S/N$_{rem}$\,$>$\,6         &   1760     &   4018  & 69.54 \\  
      Transit points\,$>$\,18         &    563     &   1197  & 68.01 \\  
      V$_{shape}$\,$<$\,0.7           &    360     &    203  & 36.06 \\  
      Depth\,$\leq$\,3\,\%            &    100     &    260  & 72.22 \\  
      Transit duration\,$\leq$\,0.5   &    100     &      0  &  0.00  \\
      \hline
    \end{tabular}
  }
  \label{tab:selec_aperture}
  \tablefoot{Number of AP candidates after applying all criteria
    selection (100).  }
\end{table}

We visually inspected the 200 detections that pass the optimized
selection criteria in the AP and DI data sets and removed candidates,
which are clear eclipsing binaries with two eclipses of different
depth, objects that show significant out of eclipse variations and
very asymmetric eclipse shapes, as well as, candidates which are too
noisy to be further analyzed. We also eliminated objects that have
periods below 0.5\,days. Our final list of candidates includes 11
objects, of which 7 were detected in the AP-light curves and 6 are
from the DI light curves. Two objects are common detections in both
the DI and AP light curves, of which one candidate is WTS-2b that has
recently been confirmed as a planet by the RoPACS community
\citep{2013EPJWC..4701004B,2013MNRAS..111..111}. WTS-1b is the other
planet that has been found in the WTS \citep{2012MNRAS.427.1877C},
which was not detected by our selection criteria due to a very low S/N
value.\\ In the following we present a detailed analysis of the 10
remaining candidates including a characterization of the host stars, a
light curve fit with an analytic transit model and a test for
double-eclipse binary scenarios. The analysis provides important
physical parameters of the host stars and companions, which are used
to asses the quality of the candidates. Figure
\ref{fig:all_candidates} shows the folded light curves of our
candidates.\\

\subsection{Characterization of the host star}
\label{sec:paren_star}

The broad band photometric measurements of the host stars of the
candidates are listed in Table \ref{tab:photometry}. The WFCAM
provides photometry in five bands (Z,Y,J,H,K). Additional measurements
in five optical bands (u,g,r,i,z) were obtained from the database of
the Sloan Digital Sky Survey (SDSS 7$^{th}$ release,
\citealt{2009yCat.2294....0A}). The table also shows in which data-set
the candidate was detected (AP or DI). The candidate 19b1-02162 was
found in both AP and DI data sets, in this case we use the AP light
curve in the following, since it presents a lower scatter.\\

\begin{table*}\bf
  \caption{List of new candidates around F-G-K stars detected in this work.}
\scalebox{1.00}{
 \centering
  \begin{tabular}{l*{16}{c}}\hline
    Object      & Data-set  & $\alpha$  & $\delta$ & u     & g     & r     & i     & z     & Z     & Y     & J     & H     & K     \\\hline
19b1-02162 &  AP/DI &  293.0112 & 36.4848 & 19.00 & 17.73 & 17.13 & 16.89 & 16.76 & 16.37 & 16.22 & 15.93 & 15.49 & 15.37\\
19f3-06991 &    AP  &  293.4682 & 36.4995 & 15.97 & 14.66 & 14.26 & 14.13 & 14.07 & 13.62 & 13.56 & 13.34 & 13.07 & 13.02\\
19b3-09004 &    DI  &  293.5208 & 36.8839 & 17.97 & 16.60 & 16.03 & 15.80 & 15.67 & 15.25 & 15.17 & 14.85 & 14.45 & 14.39\\
19g1-11212 &    AP  &  293.6753 & 36.1420 & 16.57 & 15.40 & 14.93 & 14.81 & 14.74 & 14.32 & 14.25 & 14.01 & 13.73 & 13.65\\
19c4-02952 &    DI  &  293.8666 & 36.7571 & 18.53 & 17.38 & 16.97 & 16.79 & 16.70 & 16.23 & 16.11 & 15.83 & 15.51 & 15.46\\
19h1-00325 &    AP  &  294.1531 & 36.0794 & 19.55 & 17.18 & 16.19 & 15.84 & 15.60 & 15.28 & 15.03 & 14.60 & 14.09 & 13.91\\ 
19b3-05398 &    AP  &  293.4401 & 36.7404 & 20.51 & 18.67 & 18.08 & 17.81 & 17.65 & 17.24 & 17.12 & 16.78 & 16.40 & 16.33\\
19e1-05755 &    DI  &  292.6870 & 36.2186 & 18.04 & 17.09 & 16.54 & 16.29 & 16.20 & 15.84 & 15.73 & 15.42 & 15.08 & 15.01\\
19b4-04138 &    AP  &  292.9365 & 36.7902 & 17.04 & 15.70 & 15.18 & 14.95 & 14.86 & 14.38 & 14.28 & 13.96 & 13.54 & 13.47\\
19b2-01819 &    DI  &  293.5220 & 36.4675 & 18.27 & 16.95 & 16.48 & 16.32 & 16.25 & 15.86 & 15.75 & 15.46 & 15.13 & 15.07\\\hline  
  \end{tabular}
}
  \label{tab:photometry}
  \tablefoot{The second column shows the light curve data set in which
    the candidates have been detected. The coordinates (J2000.0) are
    listed in columns 3 and 4. The remaining columns provide broad
    band photometric measurements of our candidates in ten different
    filters. The u, g, r, i and z AB-magnitudes were obtained from the
    Sloan Digital Sky Survey (SDSS) and the Z, Y, J, H, K magnitudes
    are WFCAM measurements in the Vega-system.}
\end{table*}

\begin{figure}[h!]
  \centering
    \includegraphics[width=0.4\textwidth]{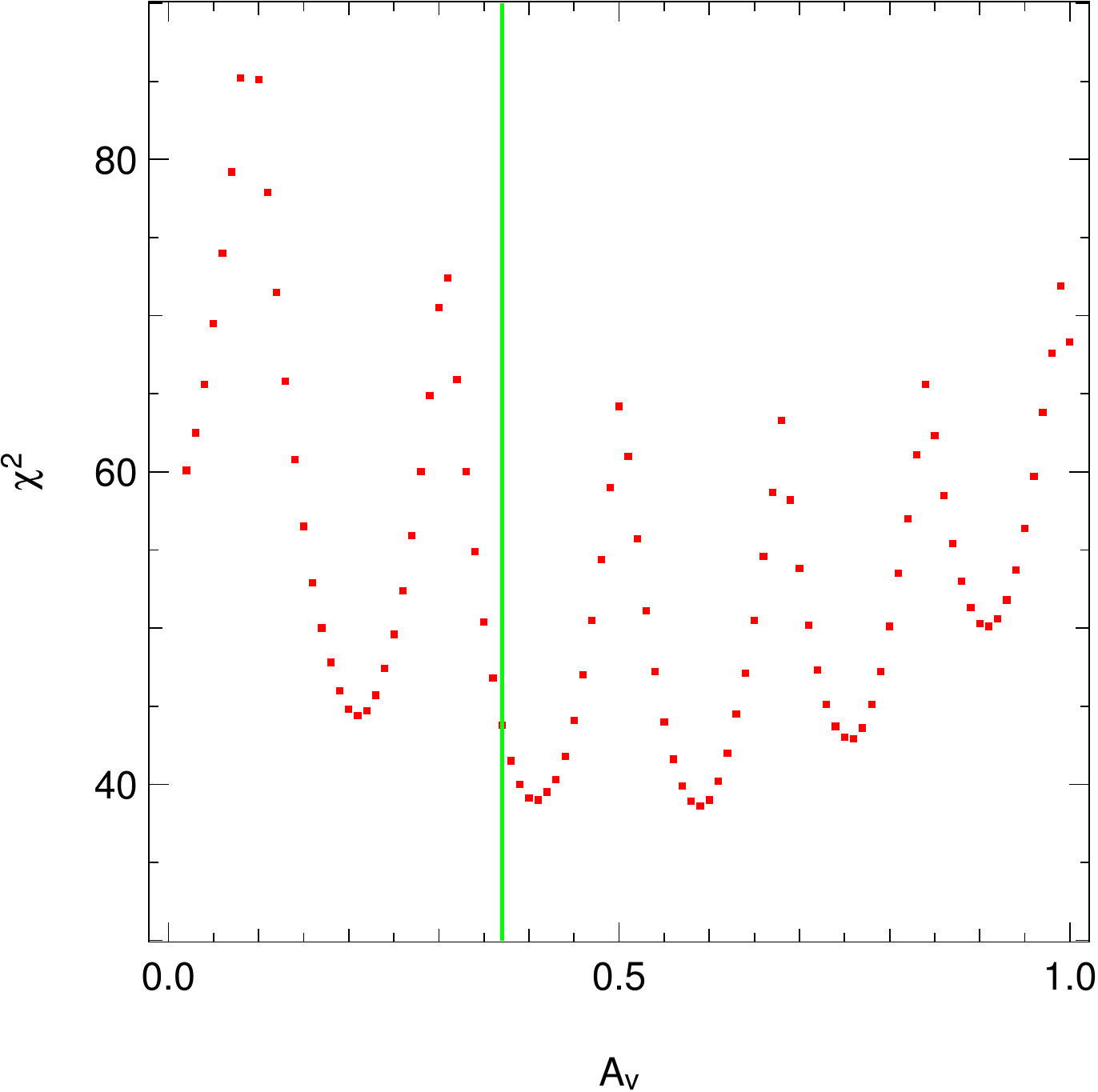}
    \caption{$\chi^2$ as a function of the input visual extinction
      value used in the SED fit of our planet candidate
      19b1-02162. Although the value of
      A$_{\mathrm{v}}$$\sim$0.6\,mag results in the lowest $\chi^2$,
      we use the value of A$_{\mathrm{v}}$=0.21\,mag based on the
      upper limit extinction adopted from the NASA/IPAC Extragalactic
      Database, since an extinction of A$_{\mathrm{v}}$=0.6\,mag would
      be physically non-realistic. The upper limit mentioned above is
      pointed out with the green solid line. The
        periodic distribution of the $\chi^2$ is due to the variation
        of six different stellar spectral-types.}
  \label{fig:19g_2_01326_points}
\end{figure}

\begin{figure}[h!]
  \centering
    \includegraphics[width=0.4\textwidth]{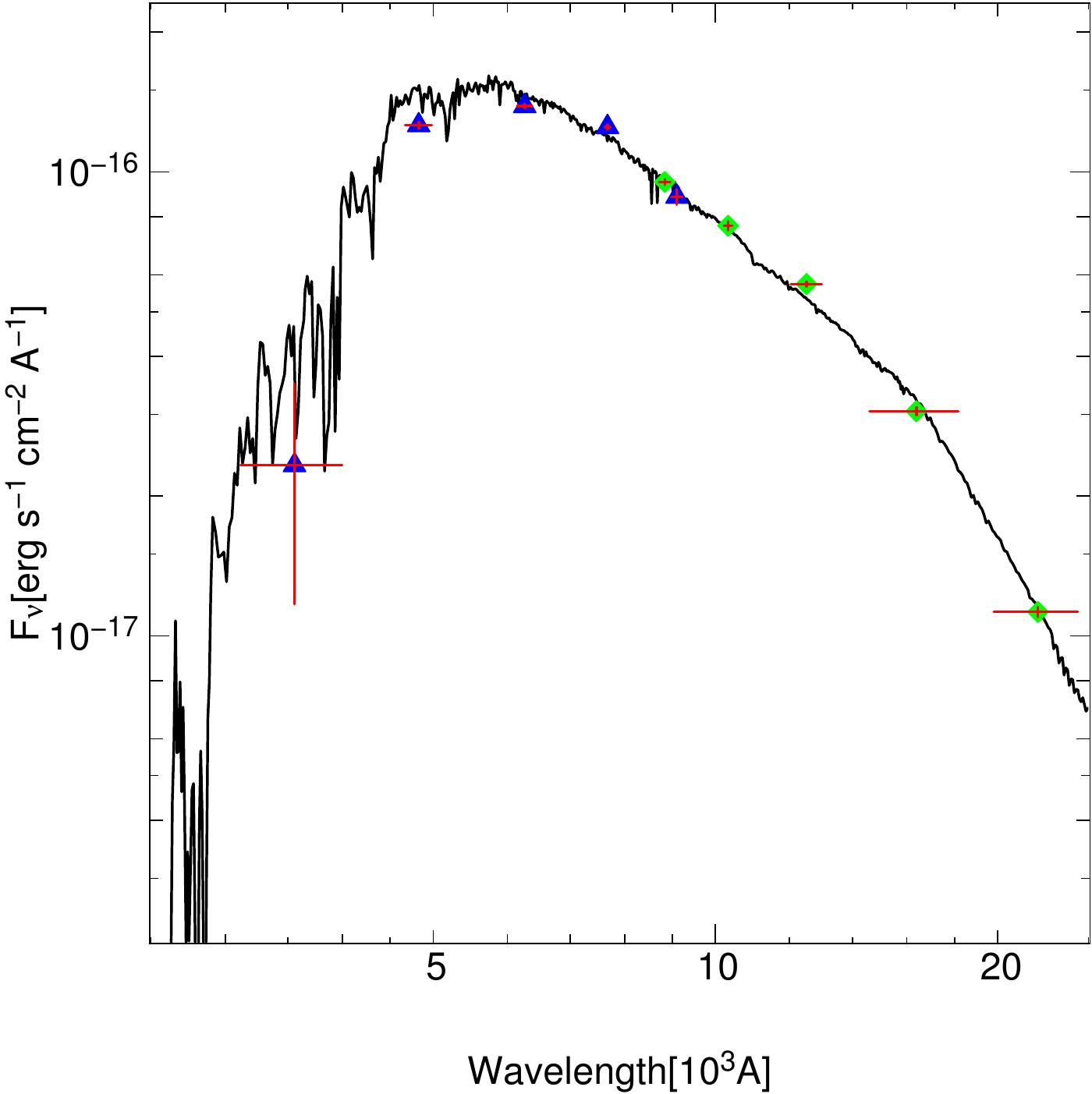}
    \caption{Best Kurucz ATLAS9 model derived with VOSA (black line)
      for the SED of 19b1-02162 . The effective temperature of the
      best fitting model is $T_{eff}$\,=\,5500\,K for an extinction of
      A$_{\mathrm{v}}$\,=\,0.21. Blue triangles represent the SDSS
      photometry while the green diamonds correspond to the WFCAM
      photometry. Vertical and horizontal errors bars are the flux
      uncertainties and the equivalent width of each pass-band.}
  \label{fig:19g_2_01326_vosa}
\end{figure}

The characterization of the host star is essential to infer physical
properties of the candidates, such as planetary radius and orbit
inclination. The Virtual Observatory SED
Analyzer\footnote{\url{http://svo2.cab.inta-csic.es/theory/vosa/}}(VOSA,
\citealt{2008A&A...492..277B}) is an on-line tool designed to
automatically perform several tasks, such as the determination of
stellar parameters by analyzing the SED. This analysis was carried out
in our candidates using the photometry reported in Table
\ref{tab:photometry}. VOSA works with input parameters that can be
submitted as ASCII files. They must include a reference name of the
source, coordinates, visual extinction A$_{\mathrm{v}}$, filter names,
observed fluxes and the corresponding errors. Although VOSA enables us
to select among 6 different fitting models, only two are appropriate
for our purpose. For the F-, G- and K-stars we adopt the Kurucz ATLAS9
templates described in \citet{1997A&A...318..841C}, which provide
better results for a wider temperature range than the NextGen model
\citep{1998A&A...337..403B}. The program offers the option of
restricting free parameters ($T_{eff}$, log\,$g$ and [Fe/H]) to speed
up the fitting process. We confine the limits to
$T_{eff}$\,=\,3\,500-10\,000K, [Fe/H]=0.0 and log\,$g$=\,3.5-5.0. Note
that the selected values of $T_{eff}$ and log\,$g$ are compatible with
main-sequence stars with spectral types between A and M. The program
compares the broad band photometric measurements to theoretical
synthetic spectra to find the best SED-fitting. VOSA tests a large
range of stellar models within the given parameter limits. The SED-fit
is also sensitive to the extinction A$_{\mathrm{v}}$, which is used as
an additional free parameter. The extinction and the corresponding SED
model are obtained by testing 100 different A$_{\mathrm{v}}$ values
distributed in a range from 0.01 to 1 magnitudes and selecting the
value that results in the lowest $\chi^2$ within the valid extinction
range from 0.01 up to the maximum allowed extinction that is set by
the total extragalactic extinction as obtained from the Galactic
Extinction Calculator of the NASA/IPAC Extragalactic
Database\footnote{\url{http://ned.ipac.caltech.edu/forms/calculator.html}}
(see Figure\,\ref{fig:19g_2_01326_points}). In some cases the absolute
minimum corresponds to an absorption that is outside the allowed
range, i.e. higher than the extragalactic value. We mark these cases
with an asterisk. The resulting best-fitting model provides an
estimate of the $T_{eff}$ of the host stars, which are summarized in
Table\,\ref{tab:vosa}. The results show that the $T_{eff}$ of the
parent stars are in the range of 4750-6500\,K which corresponds to
spectral types between K3 and F5. According to the $T_{eff}$ found in
the fit, we derive stellar radii and masses and calculate the surface
gravity log\,$g$ using 1-5\,Gyr isochrones for solar metalicity
obtained from the Dartmouth stellar evolution database
\citep{2008ApJS..178...89D}. These values are reported in
  Tables \ref{tab:vosa} and \ref{tab:vosa_mdwarf} as $R1_{\star}$,
  $M_{\star}$, and log\,$g_1$. The error ranges of the stellar
radii is determined by assuming a precision of 250\,K which is the
step size of the grid used in the VOSA
fitting. Figure\,\ref{fig:19g_2_01326_vosa} shows an example of the
VOSA fit of our best candidate 19b1-02162.\\

\begin{table*}\bf
 \caption{Characterization of host stars.}
 %\scalebox{0.80}{
    \centering
    \begin{tabular}{l*{10}{c}}
      \hline
      Object      & $T_{eff}$(K) & Spectral Type & log\,$g_1$ & log\,$g_2$ &A$_{\mathrm{v}}$ & Distance(pc) & $R1_{\star}(R_{\odot})$ & $R2_{\star}(R_{\odot})$ & $M_{\star}(M_{\odot})$\\\hline\\%[0.1cm]
19b1-02162 & 5500 & G8 & 4.56 & 4.32 & 0.21* & 2188 & 0.85$^{+0.07}_{-0.05}$ & 1.12$^{+0.07}_{-0.12}$ & 0.95$^{+0.07}_{-0.06}$\\[0.25cm]
19f3-06991 & 6500 & F5 & 4.31 & 4.05 & 0.35  & 1127 & 1.23$^{+0.20}_{-0.10}$ & 1.74$^{+0.38}_{-0.16}$ & 1.25$^{+0.13}_{-0.08}$\\[0.25cm]    
19b3-09004 & 5750 & G5 & 4.52 & 4.27 & 0.28  & 1472 & 0.92$^{+0.08}_{-0.07}$ & 1.22$^{+0.03}_{-0.04}$ & 1.02$^{+0.07}_{-0.07}$\\[0.25cm]
19g1-11212 & 6250 & F7 & 4.41 & 4.00 & 0.22* & 1330 & 1.13$^{+0.20}_{-0.12}$ & 1.78$^{+0.27}_{-0.18}$ & 1.17$^{+0.09}_{-0.08}$\\[0.25cm]
19c4-02952 & 6250 & F7 & 4.41 & 4.01 & 0.44* & 3119 & 1.13$^{+0.20}_{-0.12}$ & 1.77$^{+0.11}_{-0.07}$ & 1.17$^{+0.09}_{-0.08}$\\[0.25cm]
19h1-00325 & 4750 & K3 & 4.57 & 4.43 & 0.16  &  773 & 0.71$^{+0.03}_{-0.04}$ & 0.89$^{+0.08}_{-0.05}$ & 0.78$^{+0.04}_{-0.05}$\\[0.25cm]
19b3-05398 & 6000 & G0 & 4.47 & 4.19 & 0.45  & 4345 & 1.00$^{+0.13}_{-0.08}$ & 1.39$^{+0.08}_{-0.07}$ & 1.09$^{+0.17}_{-0.07}$\\[0.25cm]
19e1-05755 & 6000 & G0 & 4.47 & 4.12 & 0.38  & 2208 & 1.00$^{+0.13}_{-0.08}$ & 1.51$^{+0.06}_{-0.14}$ & 1.09$^{+0.17}_{-0.07}$\\[0.25cm]
19b4-04138 & 5750 & G5 & 4.52 & 3.96 & 0.45  &  506 & 0.92$^{+0.08}_{-0.07}$ & 1.74$^{+0.03}_{-0.05}$ & 1.02$^{+0.07}_{-0.07}$\\[0.25cm]
19b2-01819 & 6250 & F7 & 4.41 & 3.93 & 0.38  & 2559 & 1.13$^{+0.20}_{-0.12}$ & 1.94$^{+0.04}_{-0.05}$ & 1.17$^{+0.09}_{-0.08}$\\[0.25cm]\hline 
    \end{tabular}
  %}
  \label{tab:vosa}
  \tablefoot{The $T_{eff}$ is derived from SED-fitting. We use
    1-5\,Gyr isochrones obtained from the Dartmouth stellar evolution
    database \citep{2008ApJS..178...89D} to estimate $R1_{\star}$,
    log\,$g_1$ and $M_{\star}$. The extinction values
    (A$_{\mathrm{v}}$) found in the SED fitting are reported in column
    6. In the three cases marked with an asterisk the best fitting
    extinction is higher than the total extragalactic extinction and
    we report the extinction that corresponds to the minimum $\chi^2$
    within the allowed extinction range. The stellar radii
    $R2_{\star}$ correspond to the best fitting analytic transit model
    (see Section \ref{sec:transit_fit}). The values of
      log\,$g_2$ reported in column 5 are estimated from the stellar
      radii $R2_{\star}$, which tend to be higher than $R1_{\star}$,
      resulting in lower log\,$g_2$. The distances reported in column
    7 are estimated utilizing the extinction values found in the VOSA
    analysis, the i-band magnitudes reported in Table
    \ref{tab:photometry} and the absolute magnitudes M$_{\mathrm{i}}$
    which are obtained from the isochrones.}
\end{table*}

\subsection{Secondary eclipse fit}
\label{sec:sec_transit_fit}

For each candidate we tested the possibility that we actually detected
an eclipsing binary system with similar eclipse depths where the
primary and secondary eclipse have been folded together at half the
binary period. In order to do this test, we fold the light curve of
each candidate with double the detected period and fit a primary and
secondary eclipse which are offset by 0.5 phase units assuming a
circular orbit. Note that under this assumption, our candidate
  sample may be contaminated with eclipsing binaries in high eccentric
  orbits. However, \citet{2005ApJ...628..411D} shows that only
  $\sim$10\% of the binaries studied there with periods shorter than
  12 days have eccentricities higher than 0.1. Therefore, the possible
  contamination is low to start with. Moreover, any candidate with a
  clear deeper secondary eclipse would be rejected during our visual
  inspection. Both the primary and secondary eclipses are first
fitted with a box and subsequently re-fitted with a symmetrical
trapezoid as described in Section \ref{sec:transit_detection}. A
significant difference between the depths of the primary and secondary
eclipse indicates that the candidate could be an eclipsing binary
rather than a star with a planet. Also a comparison of the
$\chi'^2_{dof}$ of the binary fit to the $\chi^2_{dof}$ of the fit
with the planet period can indicate that the candidate is actually a
binary with similar eclipse depths.  We would like to point out that
the decision of presenting either the planet or binary periods
(Figures \ref{fig:all_candidates} and
\ref{fig:all_candidates_mdwarfs}) included a visual examination of the
folded light curves. This inspection showed that $\chi_{dof}^2$ and
eclipse depths differences cannot be used blindly for the
discrimination, since they closely depend on the number of points
during the eclipses and box-fitting parameters. Note that the
trapezium fit is only a crude model of a transit light curve and in
some cases we found that the depth estimated by our algorithm did not
reflect the true depth as one sees in the folded light curves. In
summary, the discrimination between both scenarios based on the
$\chi_{dof}^2$ and eclipse depths values is only used as a hint to
select either the planet or binary period rather than a decisive proof
of the nature of the candidate. The final decision to classify our
candidates was done case-by-case and primarily based on the best
fitting radius as found in the analytic transit fit (see Section
\ref{sec:transit_fit}). Table\,\ref{tab:sec_tran} summarizes the
results of the secondary eclipse fit analysis.\\

\begin{table*}\bf
  \caption{Comparison between the planet and binary scenario.}
  \centering
  \begin{tabular}{c c c c | c c c c c}\hline 
   Object      & $V$  & dp($\%$) & $\chi^2_{dof}$ & $\chi'^2_{dof}$ & dp$'_1(\%)$ & dp$'_2$($\%$) & $V'_1$  & $V'_2$ \\\hline
19b1-02162  &  0.25  &  2.05  &  1.3792  &  1.3438  &  2.54  &  1.37  &  0.25  &  0.00\\
19f3-06991  &  0.56  &  0.81  &  1.0087  &  0.9545  &  1.08  &  0.47  &  0.58  &  0.33\\
19b3-09004  &  0.31  &  3.07  &  4.3247  &  4.2817  &  3.44  &  2.87  &  0.54  &  0.01\\
19g1-11212  &  0.37  &  1.49  &  1.4751  &  1.4301  &  2.29  &  1.34  &  0.45  &  0.81\\
19c4-02952  &  0.57  &  4.13  &  3.3622  &  3.3617  &  3.73  &  3.86  &  0.00  &  0.53\\
19h1-00325  &  0.43  &  3.11  &  4.0514  &  4.0172  &  2.92  &  3.05  &  0.16  &  0.38\\
19b3-05398  &  0.29  &  2.66  &  0.9802  &  0.9739  &  2.91  &  2.55  &  0.48  &  0.36\\
19e1-05755  &  0.29  &  1.76  &  1.7486  &  1.7211  &  2.54  &  1.66  &  0.80  &  0.54\\
19b4-04138  &  0.64  &  2.53  &  1.7270  &  1.7041  &  2.51  &  2.36  &  0.66  &  0.62\\
19b2-01819  &  0.45  &  2.80  &  2.7123  &  2.6819  &  2.74  &  3.07  &  0.61  &  0.65\\\hline
  \end{tabular}  

 \label{tab:sec_tran}
 \tablefoot{Comparison of the eclipse shapes, eclipse depths and
   $\chi^2_{dof}$ values of the planet scenario (left side of the
   table) and binary scenario (right side of the table).}
\end{table*}

\begin{table*}\bf
  \caption{Results from the transit fit.}
 \scalebox{1.00}{
 \centering
  \begin{tabular}{l*{14}{c}}\hline
    Candidate   & Period(days) &$t_0$            & $i(^{\circ})$ & R$_{planet}$(R$_{Jup}$) & R$_{planet,min}$(R$_{Jup}$) & R$_{planet,max}$(R$_{Jup}$) & $\chi^2_{dof}$ & Classification \\\hline
19b1-02162 & 0.59862739 & 2454317.7883529 & 72.01 &   1.61 &  1.40  &  1.97 & 1.31 & P\\
19f3-06991 & 0.71482077 & 2454318.3894489 & 66.54 &   1.65 &  1.43  &  3.26 & 1.00 & B\\
19b3-09004 & 3.55921358 & 2454320.9406801 & 84.31 &   2.22 &  2.10  &  2.32 & 2.07 & B\\
19g1-11212 & 2.77301797 & 2454318.9644598 & 80.23 &   2.29 &  3.44  &  1.94 & 1.21 & B\\
19c4-02952 & 3.42965118 & 2454319.6314035 & 82.25 &   3.54 &  3.36  &  3.81 & 1.84 & B\\
19h1-00325 & 0.80767863 & 2454318.2698431 & 72.23 &   3.97 &  2.10  &  4.20 & 2.01 & B\\
19b3-05398 & 0.73369311 & 2454317.9513207 & 67.12 &   4.20 &  3.53  &  4.99 & 1.18 & B\\
19e1-05755 & 0.77250704 & 2454318.1282676 & 64.34 &   5.30 &  3.00  & 13.37 & 1.21 & B\\
19b4-04138 & 1.10663897 & 2454318.0883407 & 67.34 &   6.14 &  4.17  &  8.32 & 1.30 & B\\
19b2-01819 & 0.82989549 & 2454318.2785263 & 53.23 &  15.59 & 15.43  & 15.98 & 1.27 & B\\\hline
  \end{tabular}
}
  \label{tab:summary}
  \tablefoot{Orbital and planetary parameters derived from the
    analytic transit model fitting. Only one candidate, 19b-1-02162,
    is considered to be a planet candidate. All other candidates are
    too large and most likely transiting brown dwarfs or low-mass
    stars.}
\end{table*}

\subsection{Transit fit }

\label{sec:transit_fit}
\begin{figure}[ht!]
  \centering
  \includegraphics[width=0.35\textwidth]{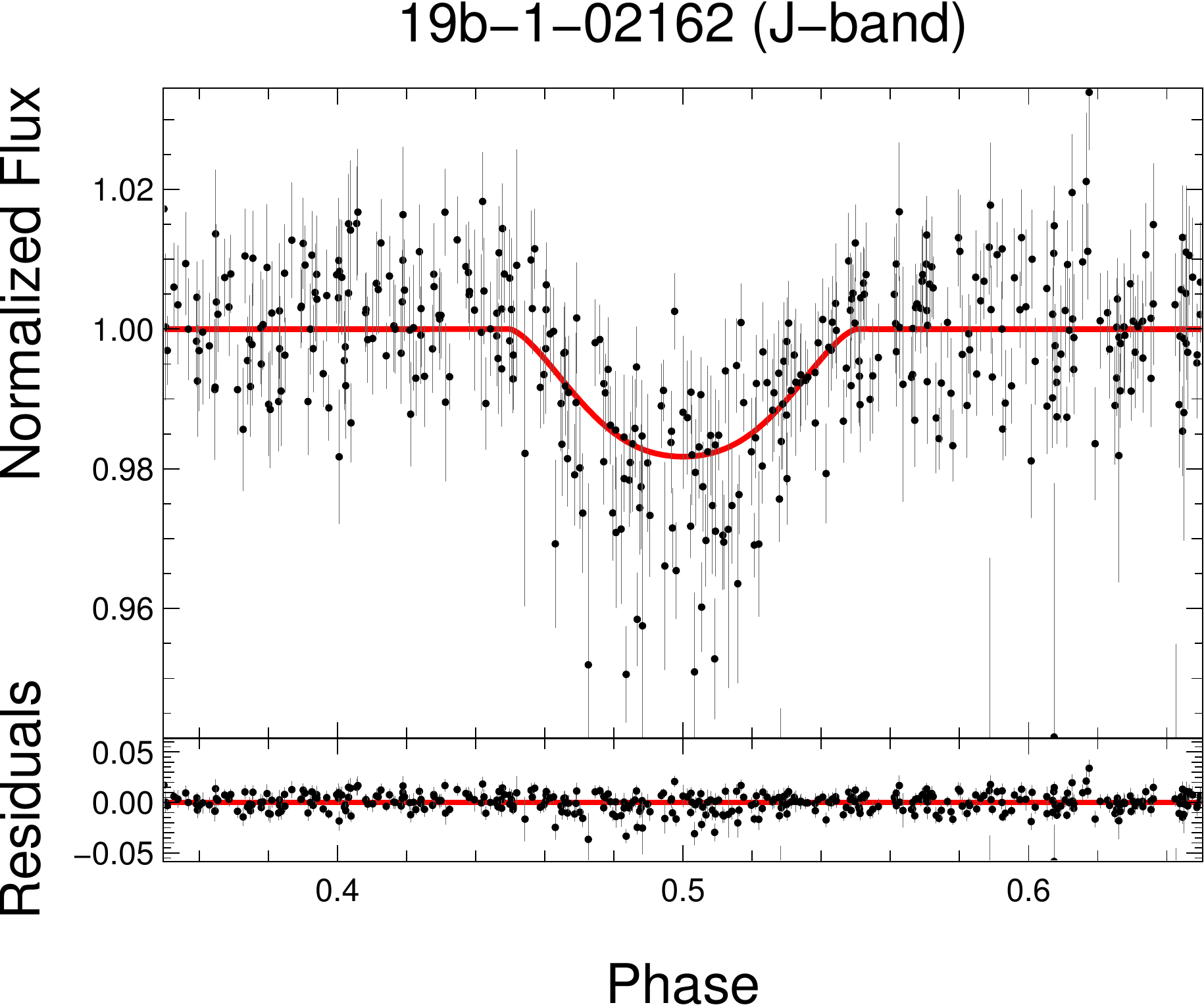}
    \caption{Best fitting model of 19b-1-02162 using the J-band
      light curve. The top frame shows the best fit, whereas the
      bottom frame represents the residuals of the fit.}
  \label{fig:19b-1-02162_fit_j}
\end{figure}

\begin{figure}[ht!]
  \centering
  \includegraphics[width=0.35\textwidth]{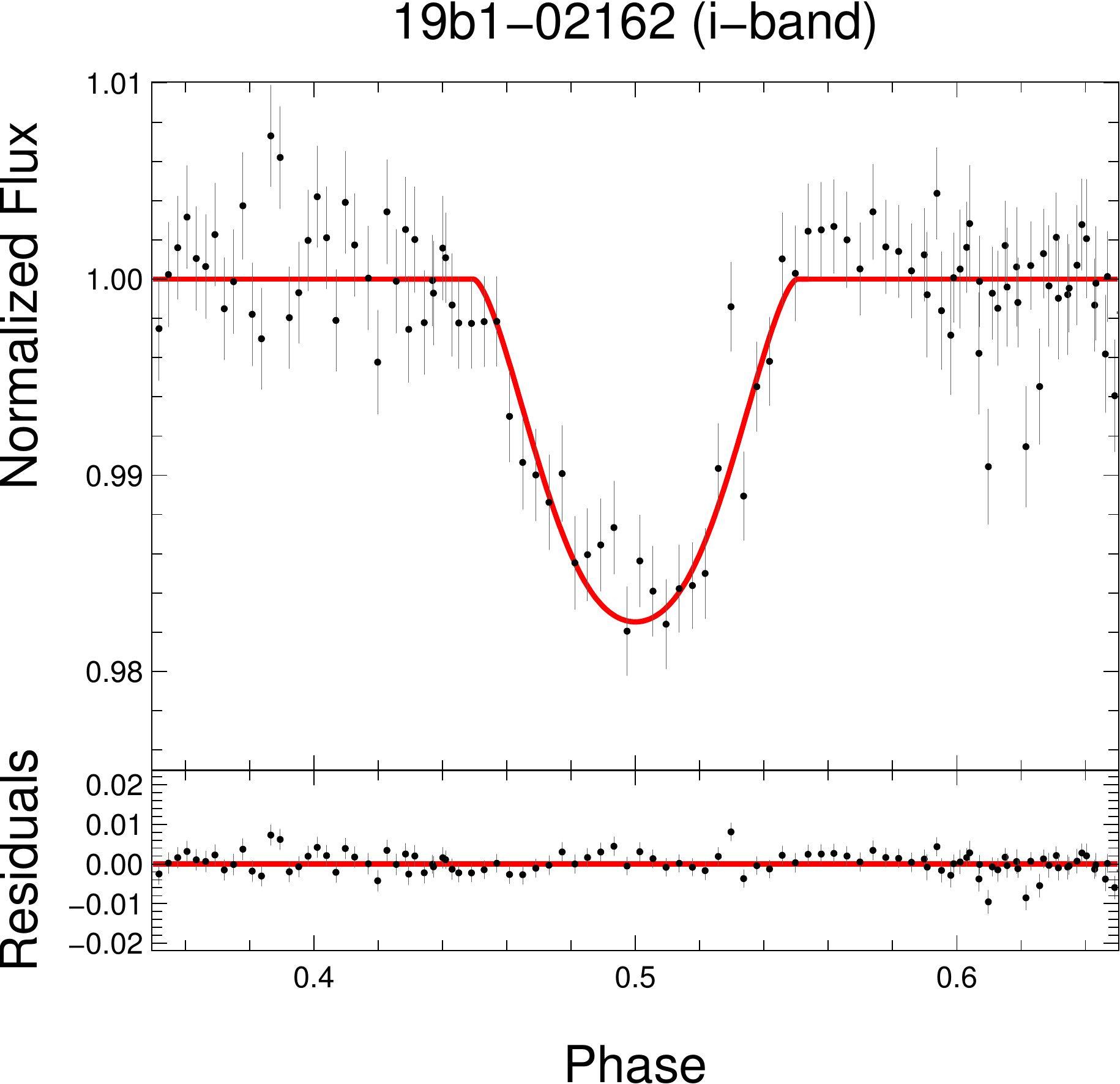}
    \caption{Best fitting model of 19b-1-02162 using the i'-band
      light curve.}
  \label{fig:19b-1-02162_fit_i}
\end{figure}

We carried out an improved fit to the J-band light curves of the
candidates using analytic transit models proposed by
\citet{2002ApJ...580L.171M}. For two candidates (19b1-02162 and
19b2-01819), we used additionally an i'-band light curve, covering one
full eclipse, which was obtained in a photometric follow-up campaign
at the Isaac Newton Telescope on La Palma. In these cases we performed
a simultaneous fit to both light curves. The transit light curve model
depends on quadratic limb-darkening coefficients, which were deduced
as linear interpolations in $T_{eff}$ and log\,$g$ of the values
listed in \citet{2011A&A...529A..75C}. We used the $T_{eff}$ of host
stars that were previously obtained by the SED analysis (see Section
\ref{sec:paren_star}) and the corresponding log\,$g$ values from the
1-5 Gyr isochrones, assuming a solar metalicity [Fe/H]=0.0 and a micro
turbulence of 2\,km/s. We utilized the values derived from ATLAS
atmospheric models using the flux conservation method
(FCM). Alternatively, the values can be derived using the
least-squares method (LSM). However, a transit fit-test using the
values from the two different models showed the same goodness of the
fit for both methods, so we have chosen the FCM over the LSM model
without any specific preference. Using the WTS J-band light curve, we
fitted the mean stellar density $\rho_{\star}\,\sim\,M_{\star}$ /
$R^{3}_{\star}$ in solar units, the radius ratio $R_{planet}$ /
$R_{\star}$, the impact parameter $\beta_{impact}$ in units of
$R_{\star}$, the orbital period $P$ and epoch of the central transit
$t_0$. The iterative fitting process required starting values for a
series of input parameters, such as period, epoch of transit, planet
radius and parameter related to the stellar companion, such as mass
and radius. The period, epoch of transit and planet radius were
obtained directly from the results provided by our transit detection
algorithm, while the stellar parameters ($R1_{\star}$ and $M_{\star}$)
were estimated by using the previously fitted $T_{eff}$ from the
1-5\,Gyr model isochrones for solar metalicity
\citep{2008ApJS..178...89D}. From the best fitting transit model, we
were able to calculate the intrinsic physical parameters of the
candidates and host stars, such as $R_{planet}$ and
$R2_{\star}$.\\ The fitting procedure also enabled us to derive an
error estimation of the fitted parameters. The errors were calculated
using a multi-dimensional grid in which we searched for extreme points
with $\Delta\chi^2$=1. This method corresponds to a variation of each
single parameter while minimizing over the others. The results of the
transit fit are listed in Table\,\ref{tab:vosa} and
\ref{tab:summary}. Figures \ref{fig:19b-1-02162_fit_j} and
\ref{fig:19b-1-02162_fit_i} show the best fitting model of our best
candidate 19b-1-02162 in the J and i'-bands respectively.\\

\subsection{Discussion of the candidates}
\label{sec:candidates_disccusion}

\begin{figure}[h!]
  \centering
    \includegraphics[width=0.3\textwidth]{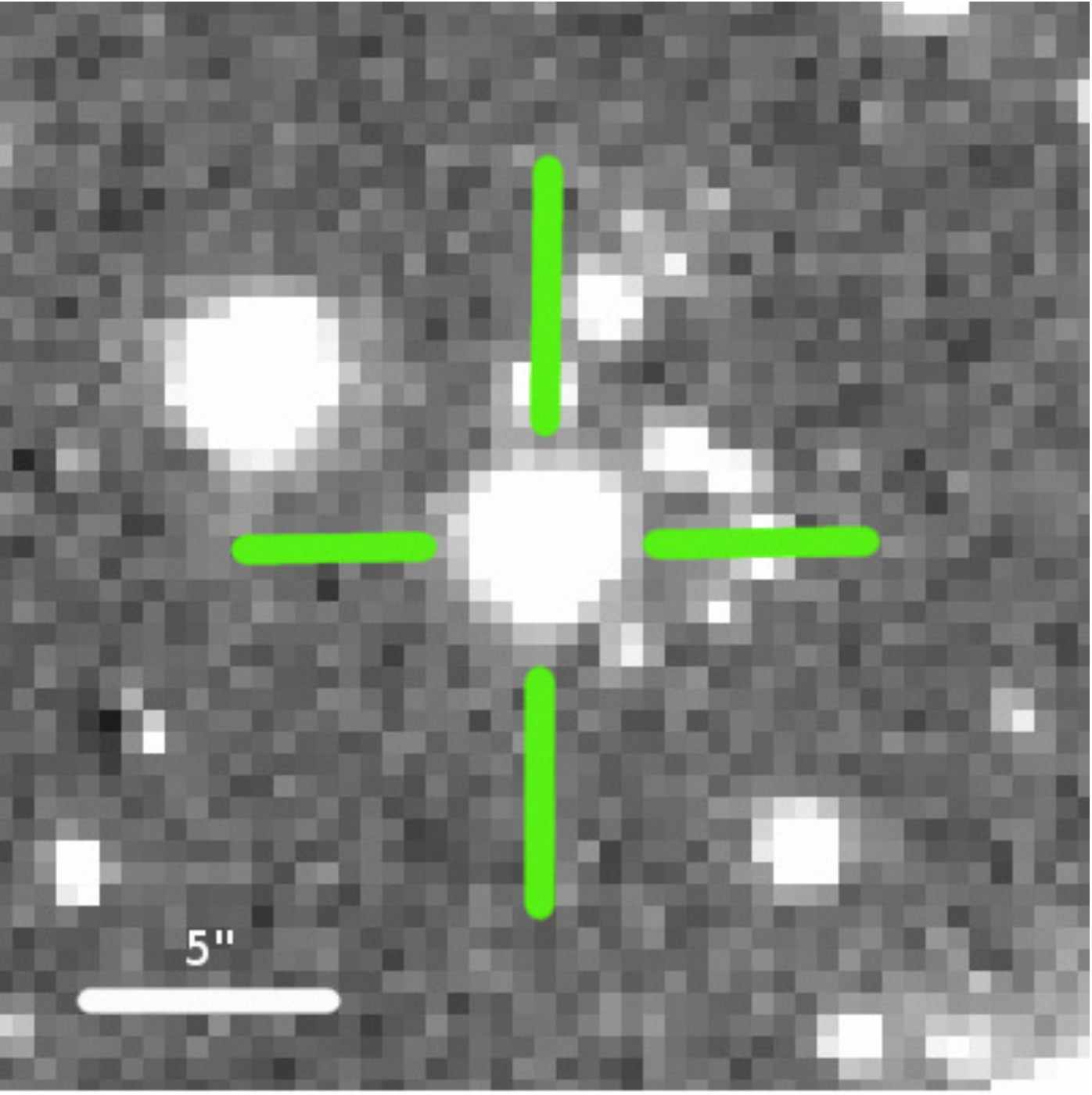}
    \caption{Zoom-in showing the crowded neighborhood of the candidate
      19b1-02162 on the sky. Our difference imaging pipeline is
      optimized to deal with such cases.}
  \label{fig:19g_2_01326_near}
\end{figure}

\begin{figure*}[!ht]
  \centering
    \includegraphics[width=1.0\textwidth]{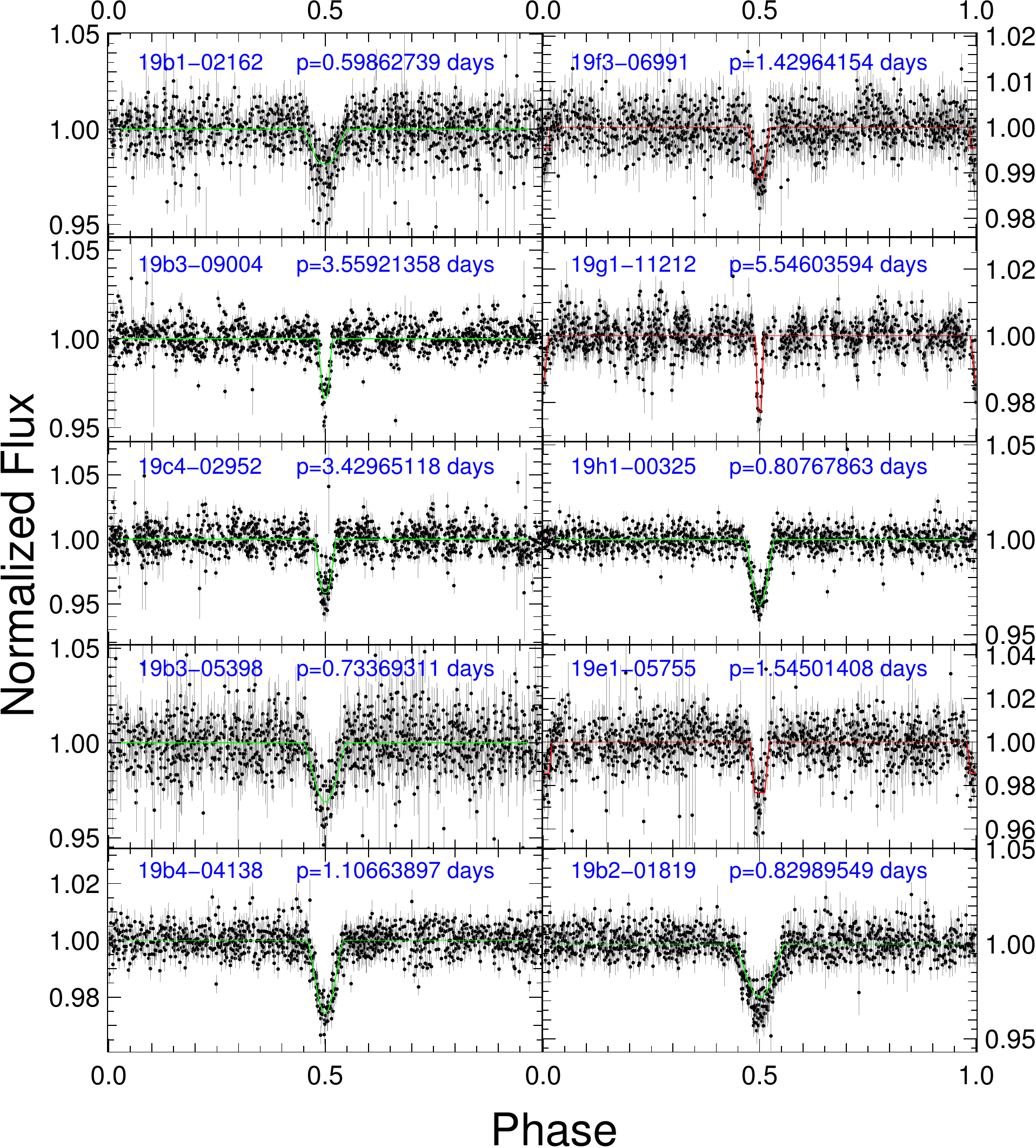}
    \caption{Phase-folded light curves of the 10 candidates orbiting
      F-G-K stars found in the WTS 19h field. We overplot the best
      fitting single eclipse transit models (green lines) for all
      objects for which the binary scenario fit does not show an
      improvement over the single eclipse scenario. The light curves
      with two eclipses are shown together with the best fitting
      trapeziod model (red lines) }
  \label{fig:all_candidates}
\end{figure*}

Table \ref{tab:summary} provides a list of our candidates sorted
according to their best fitting radius. All candidates except for the
first two have very large best fitting radii, larger than all
transiting planets published so far. We therfore conclude that they
are systems with a transiting brown dwarf or a low-mass stellar
companion.\\ The first two candidates have best fitting radii of
1.61\,R$_{Jup}$ and 1.65\,R$_{Jup}$, however, the secondary eclipse
fitting results in a slightly better $\chi^2_{dof}$ for the binary
scenario and the primary and secondary eclipse depths differ. Indeed,
looking at the folded light curves (Figure \ref{fig:all_candidates})
the second candidate, 19f3-06991, is a clear case where the fit with
the binary period reveals two well sampled eclipses with different
depths. The first candidate is not as clear. Although the binary
period fit shows two different eclipses with depths of 2.5 and
1.4\,\%, the single eclipse observed in the i'-band coincides with the
deeper eclipse but has a depth of 1.8\,\% (see Figure
\ref{fig:19b-1-02162_fit_i}), which is more close to the shallower
eclipse. We therefore conclude that for this candidate the correct
period is unclear and we propose it as a target for high precision
photometric follow-up. Figure \ref{fig:19g_2_01326_near} shows a
J-band image of 19b1-02162. \\ In order to estimate the
number of planets that we expect to find, we calculate the overall
detection efficiency in our simulations, being $\sim$1.7\,\% and
$\sim$2.4\,\% for DI and AP light curves respectively. Accounting for
an average geometrical probability of 11.9\,\% to see transits (as
derived from our Monte-Carlo simulations) and using an occurance rate
for short period Jupiter-sized planets of 0.5\,\%
\citep{2006AcA....56....1G,2012ApJS..201...15H} we estimate the number
of planets that we expect to find in the whole sample of 102\,428
light curves to be 1.0 (DI) and 1.5 (AP). This is in very good
agreement with the two planets that have been detected in the WTS so
far
\citep{2012MNRAS.427.1877C,2013EPJWC..4701004B,2013MNRAS..111..111}.\\

\section{Candidates detected around M-stars}
\label{sec:candidatesM}

\subsection{Selection criteria for M-stars:}

\begin{figure}[h!]
  \centering
    \includegraphics[width=0.35\textwidth]{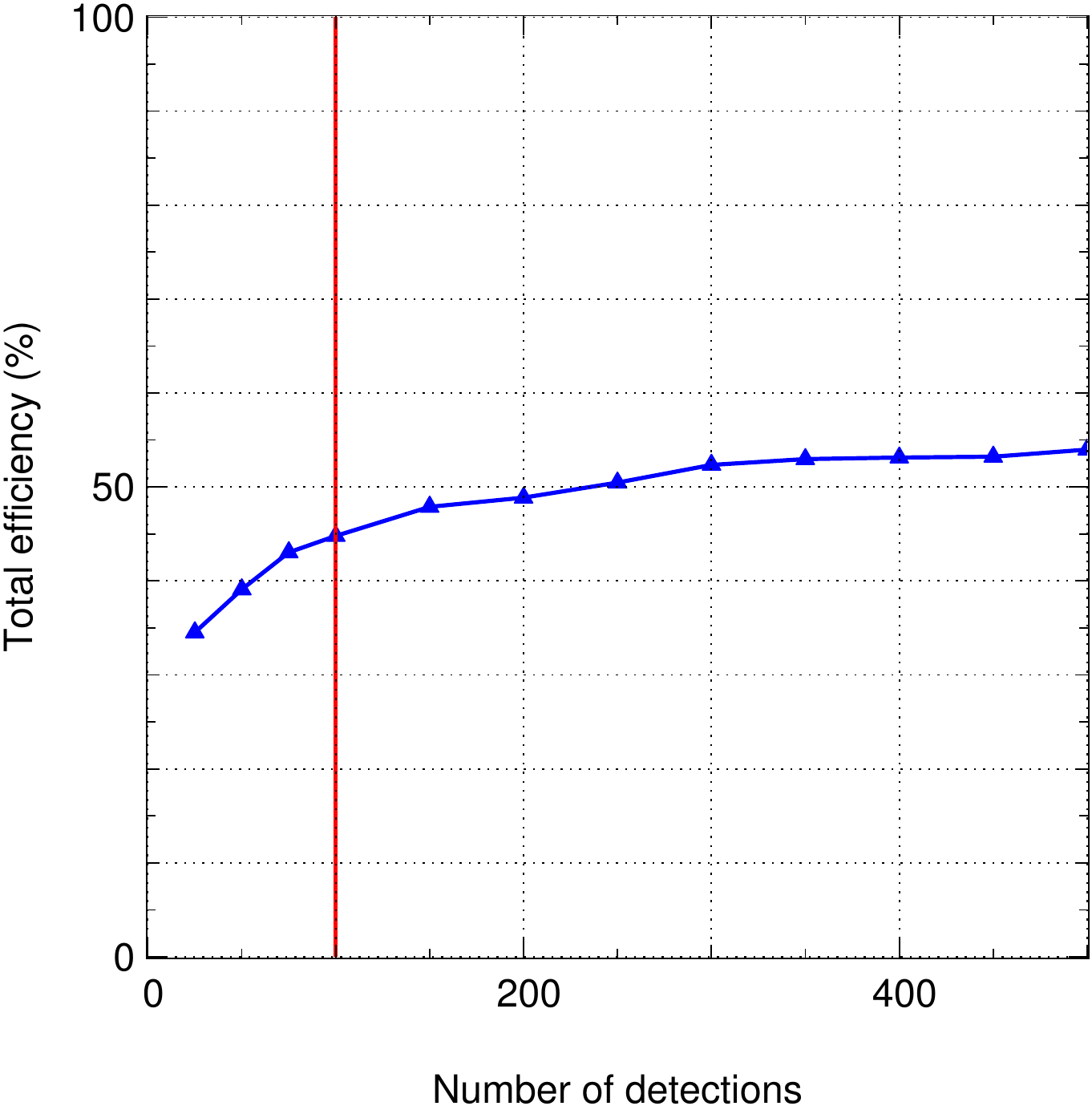}
    \caption{Optimized total detection efficiency as a function of the number of
      detections on the unmodified DI light curves. The red line shows
      our limit of 100 detections.}
  \label{fig:M_dwarf_negatives}
\end{figure}

\begin{figure}[h!]
  \centering
    \includegraphics[width=0.35\textwidth]{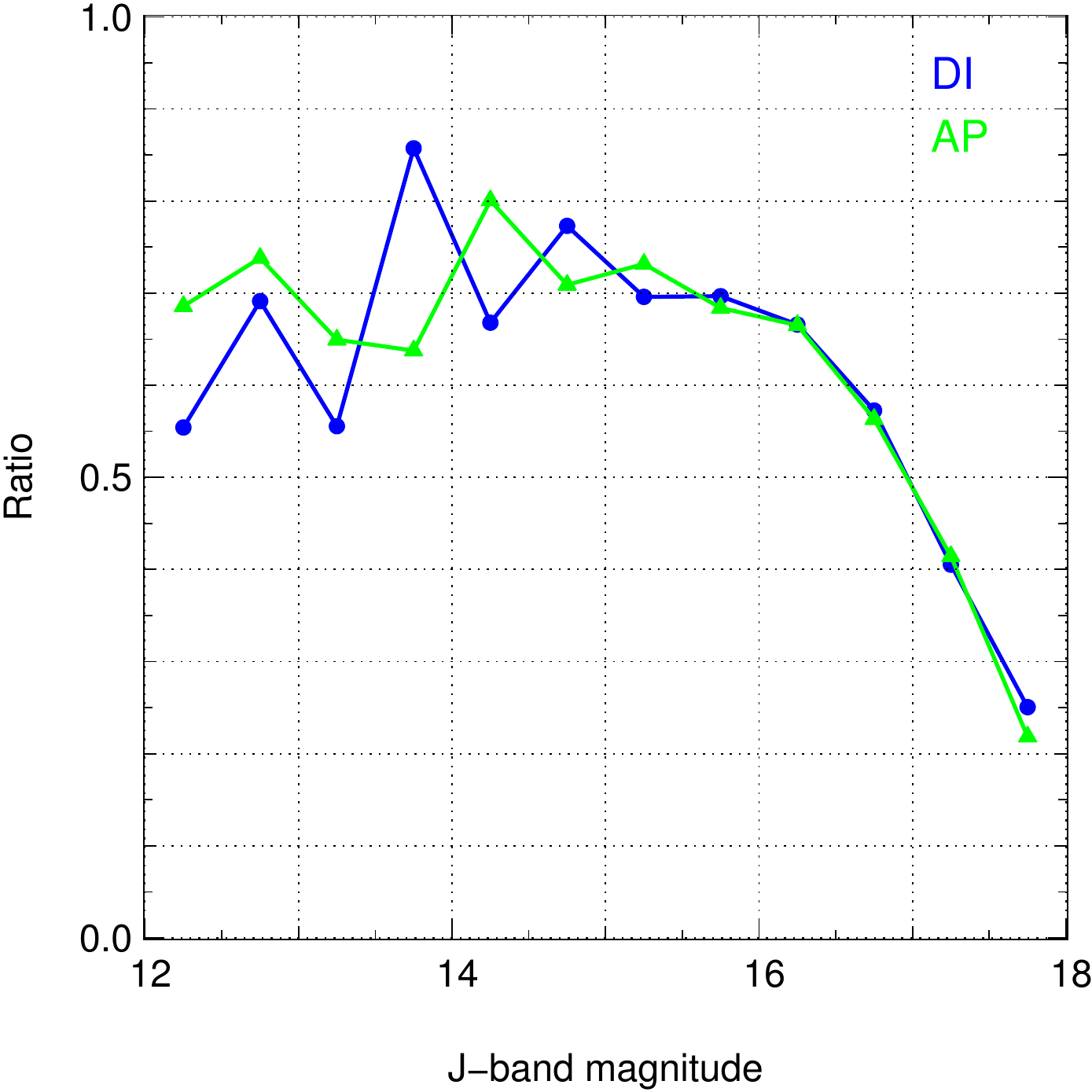}
    \caption{Detection efficiency derived from the optimized selection
      parameters as a function of the host star magnitude.}
  \label{fig:M_dwarf_efficiency}
\end{figure}

We optimized the selection criteria for M-stars by injecting
artificial transit signals into the DI and AP light curves of our WTS
M-star sample. The sample has been selected using color cuts (see
above). The simulated planet radius was always 1\,$R_{Jup}$ and we
used a flat period distribution between 0.8 and 10\,days. Using the
criteria presented above we optimized the selection of M-dwarf planet
candidates for the DI and AP light curves allowing up to 200
detections on the unmodified light curves. As for the F-, G- and
K-stars, we require the detected period differs by 1\% from the
simulated period, allowing also half or double of this value.\\ Tables
\ref{tab:selec_mdwarf_di} and \ref{tab:selec_mdwarf_ap} list the
optimized criteria for the DI and AP light curves. Unlike in the case
of F-, G- and K-stars, the fractional transit duration turned out to
be a useful selection criterion. The V$_{shape}$ parameter turned out
not to be important since transits of Jupiter-sized planets orbiting
M-dwarfs can be very V-shaped.\\ Figure \ref{fig:M_dwarf_efficiency}
shows the detection efficiency as a function of the apparent host star
magnitude. Since the total number of M-stars is dominated by the faint
end of the magnitude distribution the overall efficiency of the DI
light curves is slightly higher with 44.8\,\% with respect to 43.8\,\%
for AP light curves. Figure \ref{fig:M_dwarf_negatives} shows the
efficiency of the DI light curves as a function of the number of
detections on the unmodified light curves. Our choice of 100 provides
a high efficiency while still being managable to visually inspect.\\

\begin{table}[h!]\bf
  \caption{Optimized selection criteria for the DI M-star sample.}
   \scalebox{0.85}{
    \begin{tabular}{l*{3}{c}}
      \hline 
      Criterion                       & Remaining objects & Removed objects & \%    \\\hline    
      J\,$\leq$\,18                   &   10375           &    \ldots            &   \ldots   \\  
      Removed alias period            &    7913           &  2462           & 23.73 \\  
      S/N\,$>$ \,12                   &    1450           &  6463           & 81.68 \\  
      S/N-S/N$_{rem}$\,$>$\,5         &     536           &   914           & 63.03 \\  
      Transit points\,$>$\,8          &     164           &   372           & 69.40 \\  
      V$_{shape}$\,$<$\,1.00          &     164           &     0           &  0.00 \\  
      Depth\,$\leq$\,30\,\%           &     138           &    26           & 15.85 \\  
      Transit duration\,$\leq$\,0.06  &      98           &    40           & 28.98 \\
     \hline
    \end{tabular}
  }
  \label{tab:selec_mdwarf_di}
 % \tablefoot{Optimized selection criteria for the M-dwarf sample from the DI
 %    light curves-set.}
\end{table}

\begin{table}[h!]\bf
  \caption{Optimized selection criteria for the AP M-star sample.}
   \scalebox{0.85}{
    \begin{tabular}{l*{9}{c}}
      \hline 
      Criterion                       & Remaining objects & Removed objects & \%    \\\hline    
      J\,$\leq$\,18                   &  10375            &    \ldots            &    \ldots  \\  
      Removed alias period            &   8510            &  1865           & 17.98 \\  
      S/N\,$>$ \,6                    &   4411            &  4099           & 48.17 \\  
      S/N-S/N$_{rem}$\,$>$\,2         &    278            &  4133           & 93.70 \\  
      Transit points\,$>$\,12         &    168            &   110           & 39.57 \\  
      V$_{shape}$\,$<$\,1.0           &    168            &     0           &  0.00 \\  
      Depth\,$\leq$\,30\,\%           &    161            &     7           &  4.17 \\  
      Transit duration\,$\leq$\,0.08  &     98            &    63           & 39.13 \\ 

     \hline
    \end{tabular}
  }
  \label{tab:selec_mdwarf_ap}

\end{table}

After visual examination of the 200 automatically selected candidates
from the AP and DI light curves we identified 8 possible
candidates. All of them were found both in the AP and DI light
curves. Table \ref{tab:photometry_mdwarf} lists the coordinates and
broad-band photometric data. As for the candidates found in the F-, G-
and K-star sample we performed three different types of analyses to
further asses the possibility of them being transiting planets. The
result from the characterization of the host stars is shown in Table
\ref{tab:vosa_mdwarf}. For the M-dwarf analysis, we used the NextGen
model atmospheres, which provides a wider range for low $T_{eff}$,
being more appropriated for M-dwarfs. In this case, we restrict the
limits to $T_{eff}$\,=\,1\,800-4\,500K, [Fe/H]=0.0 and
log\,$g$=\,4.5-5.5. All 8 stars are in the range of 3300\,K $\le
T_{eff} \le$ 3900\,K corresponding to spectral types M5 to M0.\\ In
the next step we performed a comparison of the planetary with a binary
scenario corrsponding to a double-eclipse system at twice the detected
period. The results are shown in Table\ref{tab:sec_tran_mdwarf}, which
reveal for two candidates, i.e. 19a1-02980 and 19a1-10878, different
eclipse depths and a significantly better $\chi'^2$.\\ Finally, we fit
the J-band light curves of the 8 candidates with an analytic transit
model (see Section \ref{sec:transit_fit}). For all faint candidates
with J$\ge$17\,mag we used the DI light curve since the photometric
precision is higher compared to the AP light curves. For the brighter
candidates 19a1-02980 and 19a1-10878 we used the AP light curves. We
determined the best fitting period, epoch of transit, orbital
inclination and planet radius. The resulting values are listed in
Table \ref{tab:summary_mdwarf}, where the smallest of our candidates
has a radius of 2.53\,$R_{Jup}$, which exceeds the radius of any
planet previously reported. We show the folded light curves of our
candidates around M-dwarfs in Figure \ref{fig:all_candidates_mdwarfs}.

\begin{table*}\bf
  \caption{List of new candidates around M-dwarfs detected in this work.}
  \scalebox{1.00}{
    \centering
    \begin{tabular}{l*{16}{c}}\hline
      Object      & Data-set  & $\alpha$  & $\delta$ & u     & g     & r     & i     & z     & Z     & Y     & J     & H   & K  \\\hline
19b4-10711  & AP/DI & 293.0253 & 36.9168 & 24.10 & 21.85 & 20.35 & 19.07 & 18.51 & 17.99 & 17.63 & 17.17 & 16.53 & 16.25\\
19e3-01290  & AP/DI & 293.0688 & 36.5510 & 26.66 & 22.68 & 21.33 & 19.94 & 19.13 & 18.71 & 18.32 & 17.75 & 17.13 & 16.87\\
19a2-10046  & AP/DI & 293.2753 & 36.3017 & 26.61 & 23.24 & 21.08 & 20.19 & 19.34 & 18.96 & 18.57 & 17.97 & 17.35 & 17.16\\ 
19a1-02980  & AP/DI & 292.7127 & 36.3127 & 21.33 & 18.72 & 17.26 & 16.53 & 16.07 & 15.73 & 15.40 & 14.91 & 14.29 & 14.07\\
19a1-07499  & AP/DI & 292.5977 & 36.4613 & 26.39 & 21.57 & 19.97 & 18.98 & 18.46 & 18.08 & 17.68 & 17.16 & 16.54 & 16.30\\
19e3-05850  & AP/DI & 293.1396 & 36.6950 & 23.30 & 21.29 & 19.93 & 19.17 & 18.76 & 18.31 & 17.98 & 17.44 & 16.84 & 16.66\\
19a1-10878  & AP/DI & 292.5126 & 36.4273 & 22.94 & 20.63 & 19.14 & 18.10 & 17.48 & 17.20 & 16.82 & 16.29 & 15.67 & 15.40\\
19a1-01358  & AP/DI & 292.7526 & 36.4241 & 25.11 & 21.65 & 20.05 & 18.89 & 18.45 & 17.92 & 17.55 & 17.03 & 16.38 & 16.16\\\hline
    \end{tabular}
  }
  \label{tab:photometry_mdwarf}
  \tablefoot{The second column shows the light curve data set in which
    the candidates have been detected. The coordinates (J2000.0) are
    listed in columns 3 and 4. The remaining columns provide broad
    band photometric measurements of our candidates in ten different
    filters. The u, g, r, i and z AB-magnitudes were obtained from the
    Sloan Digital Sky Survey (SDSS) and the Z, Y, J, H, K magnitudes
    are WFCAM measurements in the Vega-system.}
\end{table*}

\begin{table*}\bf
 \caption{Characterization of host stars for the M-dwarfs sample.}
 %\scalebox{0.80}{
    \centering
    \begin{tabular}{l*{10}{c}}
      \hline
      Object      & $T_{eff}$(K) & Spectral Type & log\,$g_1$ & log\,$g_2$ &A$_{\mathrm{v}}$ & Distance(pc) & $R1_{\star}(R_{\odot})$ & $R2_{\star}(R_{\odot})$ & $M_{\star}(M_{\odot})$\\\hline\\%[0.1cm]
19b4-10711 & 3400 & M4 & 4.94 & 4.68 & 0.02 &  676 & 0.33$^{+0.07}_{-0.09}$  & 0.44$^{+0.01}_{-0.01}$ & 0.34$^{+0.09}_{-0.11}$ \\[0.25cm]      
19e3-01290 & 3300 & M5 & 5.02 & 4.49 & 0.18 &  703 & 0.24$^{+0.11}_{-0.09}$  & 0.45$^{+0.02}_{-0.02}$ & 0.23$^{+0.11}_{-0.10}$ \\[0.25cm]      
19a2-10046 & 3500 & M3 & 4.88 & 4.50 & 0.28*& 1554 & 0.40$^{+0.05}_{-0.07}$  & 0.61$^{+0.02}_{-0.07}$ & 0.43$^{+0.05}_{-0.09}$ \\[0.25cm]      
19a1-02980 & 3900 & M0 & 4.70 & 4.21 & 0.15 &  522 & 0.55$^{+0.03}_{-0.03}$  & 0.99$^{+0.04}_{-0.05}$ & 0.58$^{+0.04}_{-0.02}$ \\[0.25cm]      
19a1-07499 & 3600 & M2 & 4.82 & 4.42 & 0.18 & 1063 & 0.45$^{+0.03}_{-0.05}$  & 0.71$^{+0.04}_{-0.06}$ & 0.48$^{+0.04}_{-0.05}$ \\[0.25cm]      
19e3-05850 & 3800 & M1 & 4.77 & 3.98 & 0.01 & 1572 & 0.52$^{+0.03}_{-0.04}$  & 1.26$^{+0.15}_{-0.02}$ & 0.56$^{+0.02}_{-0.04}$ \\[0.25cm]
19a1-10878 & 3600 & M2 & 4.82 & 4.48 & 0.12 &  704 & 0.45$^{+0.03}_{-0.05}$  & 0.66$^{+0.01}_{-0.01}$ & 0.48$^{+0.04}_{-0.05}$ \\[0.25cm]      
19a1-01358 & 3500 & M3 & 4.88 & 4.89 & 0.11 &  844 & 0.40$^{+0.05}_{-0.07}$  & 0.39$^{+0.09}_{-0.03}$ & 0.43$^{+0.05}_{-0.09}$ \\[0.25cm]\hline    
    \end{tabular}
    %}
    \label{tab:vosa_mdwarf}
    \tablefoot{The $T_{eff}$ is derived from SED-fitting
      with VOSA. The stellar radii $R2_{\star}$ correspond to the best
      fitting analytic transit model (see Section
      \ref{sec:transit_fit}). The distances reported in column 7 are
      estimated utilizing the extinction values found in the SED
      analysis, the i-band magnitudes reported in Table
      \ref{tab:photometry} and the absolute magnitudes
      M$_{\mathrm{i}}$ which are obtained from the isochrones.}
\end{table*}

\begin{table*}\bf
  \caption{Comparison between the planet and binary scenario for candidates around M-dwarf.}
  \centering
  \begin{tabular}{c c c c | c c c c c}\hline 
   Object      & $V$  & dp($\%$) & $\chi^2_{dof}$ & $\chi'^2_{dof}$ & dp$'_1(\%)$ & dp$'_2$($\%$) & $V'_1$  & $V'_2$ \\\hline
19b4-10711  &  0.69  &  23.70   &  1.7020  &  1.6952  &  23.87  &  25.90  &  0.86  &  0.66\\
19e3-01290  &  0.66  &  24.98   &  1.0759  &  1.0715  &  25.52  &  19.95  &  0.71  &  0.52\\
19a2-10046  &  0.61  &  19.63   &  2.0426  &  2.0596  &  19.52  &  17.54  &  0.62  &  0.64\\
19a1-02980  &  0.72  &   2.17   &  1.7077  &  1.6537  &   2.70  &   1.42  &  0.71  &  0.79\\
19a1-07499  &  0.68  &   7.38   &  1.3803  &  1.3759  &   7.10  &   8.01  &  0.60  &  0.90\\
19e3-05850  &  0.00  &   7.85   &  1.1237  &  1.1203  &   7.48  &   7.42  &  0.00  &  0.16\\
19a1-10878  &  0.65  &  22.70   &  1.9194  &  1.7947  &  24.19  &  19.06  &  0.65  &  0.80\\
19a1-01358  &  0.91  &   4.67   &  1.3165  &  1.3142  &   5.28  &   3.30  &  0.66  &  0.21\\\hline
  \end{tabular}  
  
  \label{tab:sec_tran_mdwarf}
  \tablefoot{Comparison of the eclipse shapes, eclipse depths and
    $\chi^2_{dof}$ values of the planet scenario (left side of the
    table) and binary scenario (right side of the table).}
\end{table*}

\begin{table*}\bf
  \caption{Characterization of canidates around M-dwarfs according the analytic transit fit.}
 \scalebox{1.00}{
 \centering
  \begin{tabular}{l*{14}{c}}\hline
    Candidate   & Period(days) &$t_0$            & $i(^{\circ})$ & R$_{planet}$(R$_{Jup}$) & R$_{planet,min}$(R$_{Jup}$) & R$_{planet,max}$(R$_{Jup}$) & $\chi^2_{dof}$ & Classification \\\hline
19b4-10711  & 1.55274390  & 2454318.1664350 & 85.16 & 2.53 & 2.34 & 2.84  & 1.28 & B\\      
19e3-01290  & 2.46752082  & 2454318.6477813 & 85.86 & 2.64 & 2.34 & 3.20  & 1.01 & B\\      
19a2-10046  & 1.45677364  & 2454318.7173035 & 84.48 & 2.78 & 2.46 & 5.31  & 1.43 & B\\      
19a1-02980  & 1.05176697  & 2454318.6516446 & 72.80 & 3.25 & 1.93 & 6.95  & 1.31 & B\\      
19a1-07499  & 1.96038974  & 2454318.5897989 & 81.29 & 3.45 & 2.14 & 7.28  & 1.18 & B\\      
19e3-05850  & 9.20198442  & 2454320.1712614 & 86.81 & 3.48 & 3.30 & 5.03  & 1.06 & B\\ 
19a1-10878  & 1.55498531  & 2454317.9578553 & 83.43 & 3.71 & 3.49 & 3.99  & 1.32 & B\\      
19a1-01358  & 1.10712079  & 2454318.6005745 & 76.76 & 4.94 & 1.71 & 7.12  & 1.09 & B\\\hline
  \end{tabular}
 }
 \label{tab:summary_mdwarf}
 \tablefoot{Orbital and planetary parameters derived from the
   analytic transit model fitting. All candidates are too large and
   are most likely transiting brown dwarfs or low-mass stars.}
\end{table*}

Since none of our candidates has a best fitting radius in the
planetary regime we conclude that they are all transiting brown dwarfs
or low-mass stars and we therefore confirm the hypothesis presented in
\citet{2013MNRAS.tmp.1446K} about the null detection of Jupiter-sized
planets around M-dwarfs in the WTS. Following their approach we
derived a 95\,\% confidence upper limit on the giant planet occurrence
rate for M-dwarfs. \citet{2013MNRAS.tmp.1446K} analyzed all sources
with J\,$\leq$\,17\,mag and found an upper limit of 1.7-2.0\,\% for
M0-M4 spectral types. In this work, we extended the search to all
M-type stars with J\,$\leq$\,18\,mag. The extra magnitude bin
increased the number of sources by a factor of 2.8. In addition, we
introduced an automatic selection procedure that reduces the number of
candidates to be visually inspected to 100 for each set of light
curves.\\ Assuming none of the candidates presented above are planets,
we set an upper limit on the giant planet occurrence rate. Using
equation (6) of \citet{2013MNRAS.tmp.1446K} and the overall detection
efficiency of 44.8\,\% for DI light curves, the average geometrical
probability to see eclipses and the total number of sources of
10\,375, the resulting upper limit is 1.1\,\%.\\

\section{ Other applications of the WTS DI light curves}
\label{sec:applications}

In the previous sections we discussed the benefits of using the WTS DI
light curves to detect transiting planet candidates, particularly when
searching for objects with faint magnitudes (J\,$>$\,16). The DI light
curves can be used for additional analyses, such as detection and
characterization of faint variable stars. In the next sections we
describe two examples of the results presented by
\citet{2012MNRAS.425..950N} and \citet{2012MNRAS.426.1507B}, which
describe the discovery of extremely-short period M-dwarf eclipsing
binaries and M-dwarf eclipsing binaries (MEBs) in the WTS. The
motivation of showing these cases is to demonstrate that DI light
curves are able to improve the results reported in the literature and
provide new eclipsing binaries candidates when extending the search to
fainter magnitudes.\\

\subsection{Extremely-short period eclipsing binaries}
\label{sec:extra_short}

\begin{figure*}[ht!]
  \centering
    \includegraphics[width=1.0\textwidth]{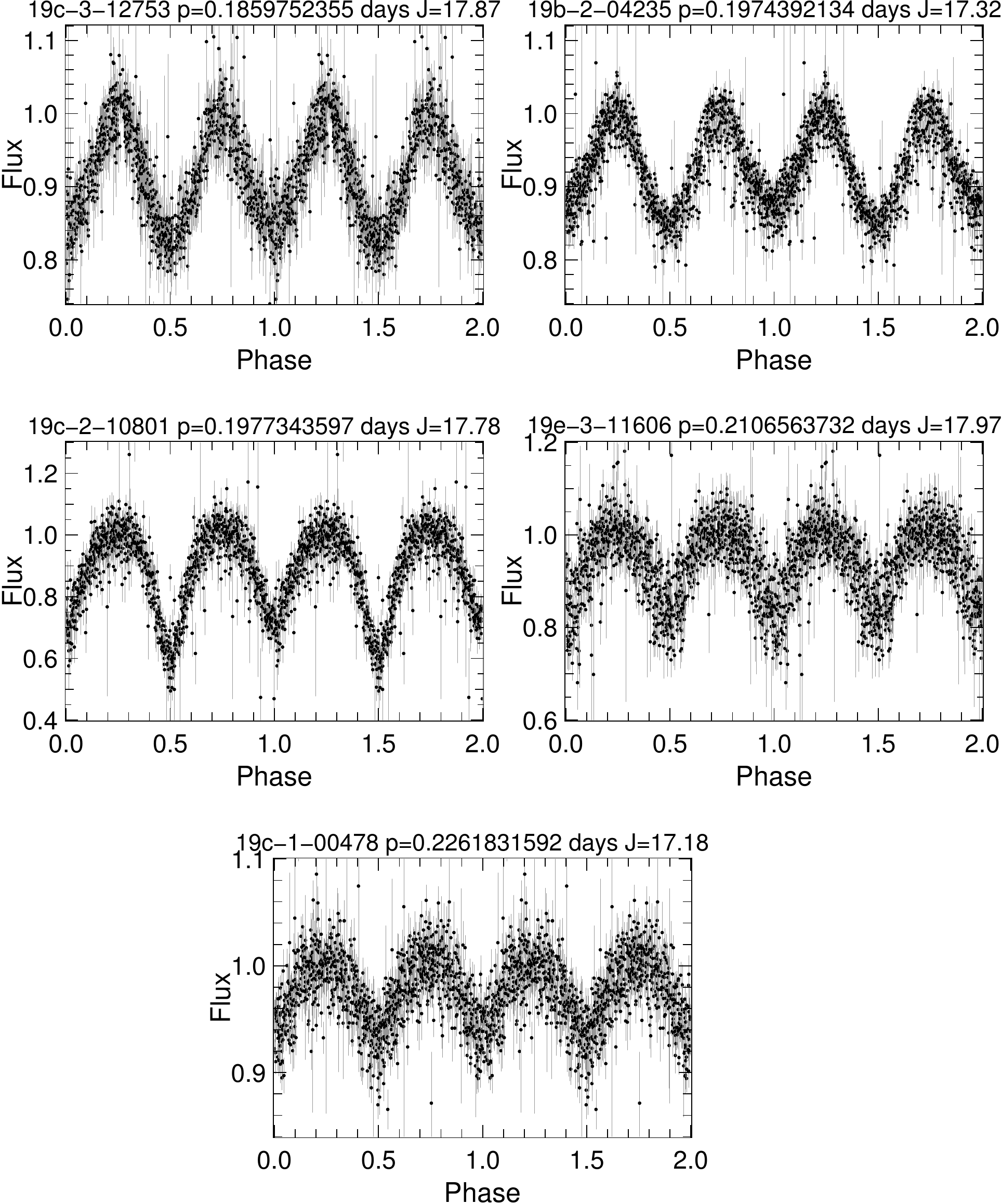}
    \caption{Phase-folded light curves of the five extremely-short
      period eclipsing binaries found in this work.}
  \label{fig:all_candidates_speb}
\end{figure*}

Eclipsing binary stars with extremely-short periods below
$\sim$\,0.22\,days are very rare systems
\citep{1992AJ....103..960R,2011A&A...528A..90N}. So far, only a few of
such objects have been discovered (e.g. \citealt{2010MNRAS.406.2559D};
\citealt{2004A&A...426..577M}). The parameters of these systems can
put strong constraints on formation and evolution theories of low-mass
stars \citep{2005ApJ...628..411D,2007ApJ...663..249D}. Recently,
\citet{2012MNRAS.425..950N} reported a sample of 31 eclipsing binaries
with periods smaller than 0.3\,days found in the WTS 03, 07, 17 and
19h fields. Four of them are M-dwarf binaries with orbital periods
considerably shorter than the sharp cut-off period of $\sim$\,0.22
days. We ran our detection algorithm on the DI light curves using the
same input parameters reported in \citet{2012MNRAS.425..950N}. We
reproduce periods and t$_0$ values of the objects reported
previously. In addition, we detected five new eclipsing binaries with
periods shorter than 0.23\,days. All systems satisfy the color cuts
and fit with the red sample (i.e M-dwarfs) presented in
\citet{2012MNRAS.425..950N}. We additionally use the SDSS color
criteria from \citet{2005AJ....129.1096I} to eliminate the possibility
of being in presence of RR Lyrae. Another cases of false-positives are
caused by contamination effects from nearby stars and stellar
variability originated by star spots. However, we reject both
scenarios, since DI method is designed to reduce the effects produced
by very near stellar neighbors and the phase folded light curves do
not present a large scatter in their amplitude, which is generally an
indication of variability generated by star spots.  Table
\ref{tab:table_speb} lists the parameters of these systems. Note that
for 19c-2-10801, the periodic signal could not be found in the AP
light curve at all. We checked for a missmatch in the
cross-identification procedure but could not find any object in the
vicinity with a comparable variability. We show the folded light
curves of all five objects in Figure \ref{fig:all_candidates_speb}.

\begin{figure}[h]
  \centering
    \includegraphics[width=0.50\textwidth]{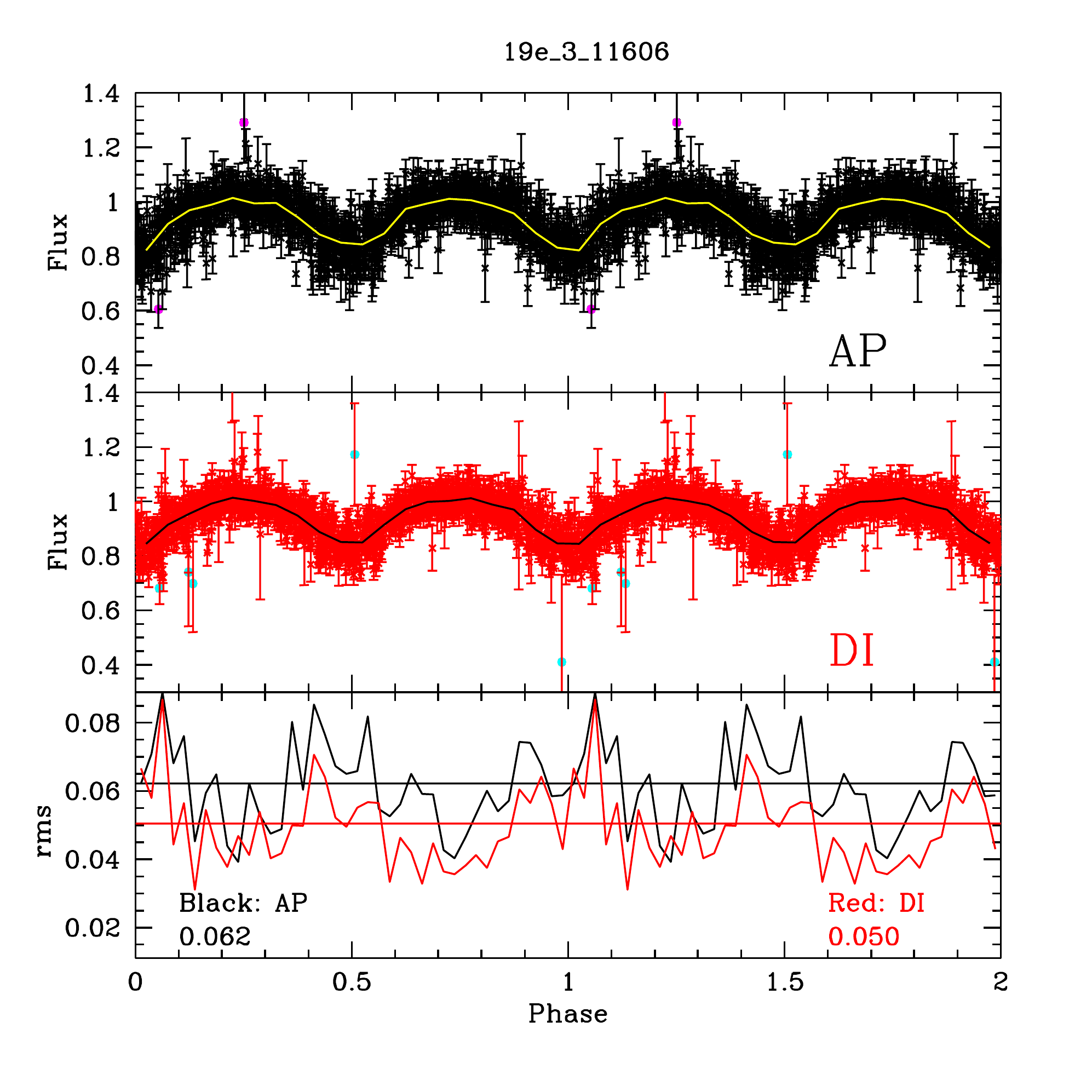}
    \caption{Comparison between the AP and DI light curve of an
      extremely-short period eclipsing binary system (19e-3-11606)
      found with the DI light curves. Black and red points in the
      uppert two panels correspond to the AP and DI light curves
      respectively. The yellow and black solid lines connect the
      median values in 40 bins with a size of 0.025 phase units. The
      lower panel shows the 4$\sigma$ clipped RMS of the residuals in
      each bin after subtracting the median. The horizontal lines
      represent the average RMS of the residuals which are 0.062 and
      0.050 for the AP and DI light curve respectively.}
  \label{fig:speb}
\end{figure}

In order to show the improvement in the precision of the DI light
curves at faint magnitudes, we carry out a statistical comparison of
the AP and DI light curves for the system 19e-3-11606, which has a
brightness of J=17.97\,mag. Figure \ref{fig:speb} shows the
phase-folded DI and AP light curves and the RMS with respect to the
mean in 40 equally spaced bins. The horizontal lines show the
4$\sigma$ clipped RMS for both light curves which are 0.050 and 0.062,
respectively. The DI light curve therefore has an RMS that is about
12\,mmag lower, being a little bit less than what we expected for a
J=18.0\,mag object. Looking at the other four detected objects we find
that this is a general trend. The lower difference in RMS can be
explained by the fact that we are looking at variable objects for
which the $sysrem$ algorithm cannot reduce systematic effects in an
efficient way. It seems that the AP light curves are less affected by
this than the DI light curves.\\

\begin{table*}\bf
  \caption{List of extremely short period eclipsing binary systems found in this work.}
  \scalebox{0.78}{
    \centering
    \begin{tabular}{cccccccccccccccc}
      \hline
      Object      & $\alpha$ & $\delta$ & Period(days) & $t_0$           & dp$'_2$/dp$'_1$ & J     & u     & g     & r     & i     & z     & (r-i) & (i-z) & RMS(AP) & RMS(DI)  \\\hline
      19c3-12753 & 294.3839 & 36.9062  & 0.1859752355 & 2454317.8449795 & 1.56 & 17.87 & 24.93 & 22.80 & 21.09 & 20.08 & 19.17 & 1.01  & 0.91  & 0.041   & 0.028    \\
      19b2-04235 & 293.3342 & 36.4255  & 0.1974392134 & 2454317.9485581 & 1.52 & 17.32 & 22.68 & 22.07 & 20.16 & 19.43 & 18.77 & 0.74  & 0.66  & 0.027   & 0.023    \\
      19c2-10801 & 294.2404 & 36.3471  & 0.1977343597 & 2454317.9436689 & 0.88 & 17.78 & 25.86 & 21.37 & 20.24 & 19.66 & 19.36 & 0.59  & 0.29  & \ldots  & 0.049    \\
      19e3-11606 & 293.2310 & 36.6396  & 0.2106563732 & 2454317.7894282 & 1.45 & 17.97 & 25.01 & 21.71 & 20.31 & 19.67 & 19.40 & 0.65  & 0.26  & 0.062   & 0.050    \\
      19c1-00478 & 293.8732 & 36.4661  & 0.2261831592 & 2454317.8755733 & 1.57 & 17.18 & 23.53 & 20.44 & 19.40 & 18.70 & 18.35 & 0.69  & 0.36  & 0.028   & 0.022    \\ \hline    
    \end{tabular}
    }   
    \label{tab:table_speb}
    \tablefoot{Extremely short period eclipsing binary systems with
      period below 0.23\,days. We list the period, epoch and eclipse
      depth ratio as well as the WFCAM J-band and SDSS ugriz
      photometry for each candidate. The last two columns provide
      information on the 4$\sigma$ clipped RMS of the light curves
      after removing the periodic signal. In general the precision of
      the DI light curves is significantly better than the precision
      of the AP light curves. This is due to the fact that all objects
      are fainter than 17\,mag in the J-band, which is in the regime
      where DI light curves present an improvement over AP light
      curves (see Section \ref{sec:comparison}).}
\end{table*}

\subsection{M-dwarf eclipsing binaries}
\label{sec:meb}

\begin{figure*}[ht!]
  \centering
    \includegraphics[width=1.0\textwidth]{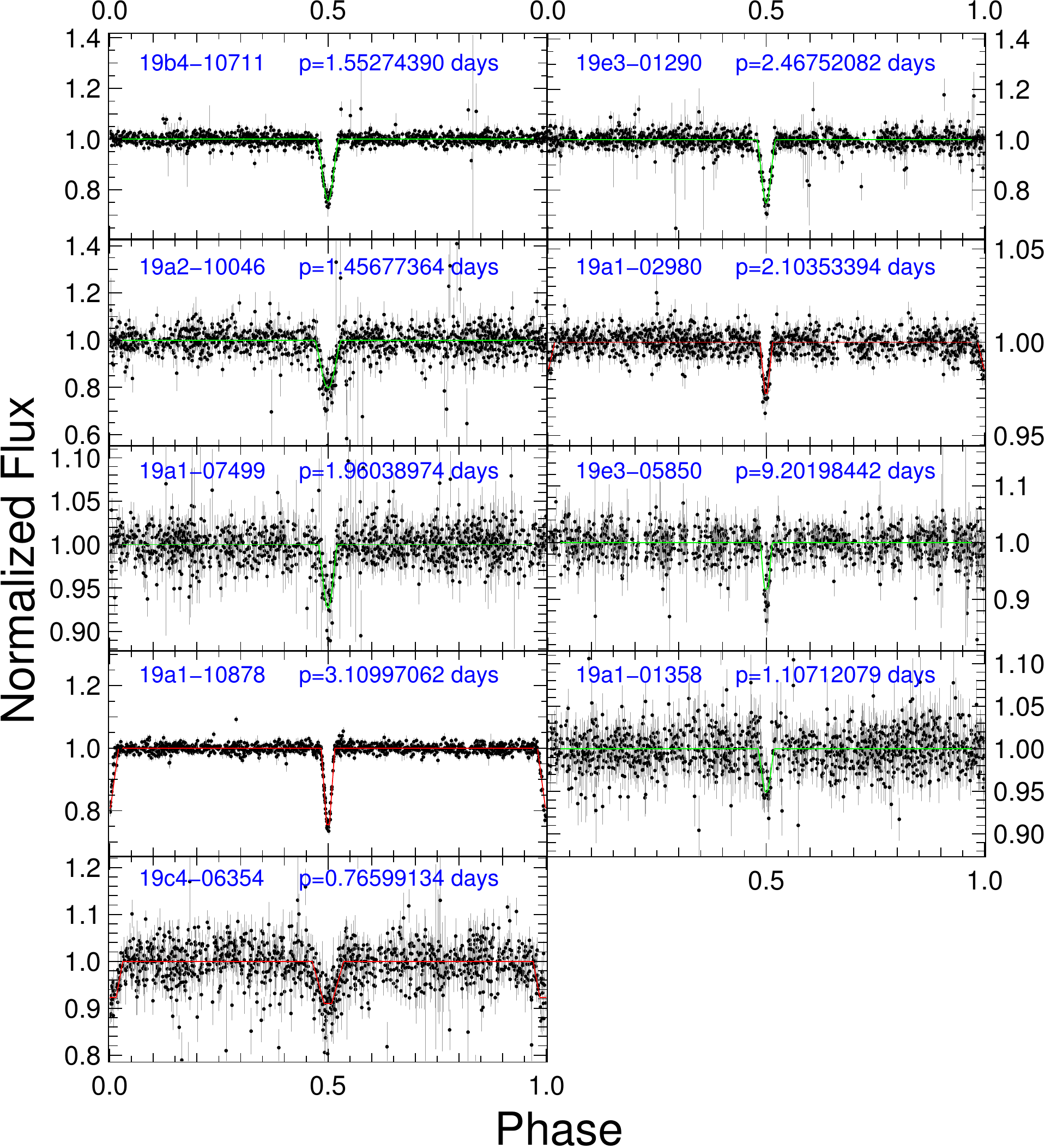}
    \caption{Phase-folded light curves of the 8 candidates orbiting
      M-dwarfs found in the WTS 19h field. As it was done for the
      candidates around F-G-K-stars, we overplot the best fitting
      single eclipse with green-solid lines, whereas the light curves
      with two eclipses are shown together with the best fitting
      trapeziod model in red-solid lines. The objects with two
      eclipses are reported in Section \ref{sec:meb} as MEB
      candidates. The objects 19a1-02980 and 19a1-10878 were found by
      our selection criteria during the process of transiting planet
      detection, whereas the system 19c4-06354 was separately detected
      by an additional analysis carried out on the AP and DI light
      curves.}
  \label{fig:all_candidates_mdwarfs}
\end{figure*}

Recently, \citet{2012MNRAS.426.1507B} reported the detection of 16
M-dwarf eclipsing binary systems (MEBs) with J$<$16 mag found in the
WTS AP light curves. These systems are particularly interesting
because they provide important information about the fundamental
properties of the most abundant stars in our Galaxy
\citep{1997AJ....114..388H}. Nevertheless, the existing theoretical
models that describe the evolution of low-mass stars differ from the
observed properties of M-dwarfs \citep{2005ApJ...631.1120L}. More
observations and characterization of MEBs can provide new evidence to
develop better and more accurate low-mass stellar evolution models
\citep{2012MNRAS.426.1507B}. We investigate the potential of extending
the search for MEBs to fainter systems with magnitudes
J\,$\leq$\,18\,mag, making use of the improvement in the photometric
precision of the DI light curves. In Table\,\ref{tab:summary_mdwarf}
we report eight candidates classified as eclipsing binary systems,
where the objects 19a1-02980 \& 19a1-10878 show strong evidence of
being MEBs. The system 19a1-02980 was actually reported and
  confirmed as MEB in \citet{2012MNRAS.426.1507B}, which supports the
  remaining fainter detections, since they were identified through the
  same process. Furthermore, we found by an additional analysis
carried out on the AP and DI light curves, a third system (19c4-06354)
with similar characteristics as the two candidates mentioned above, so
we also classify this object as MEB candidate. In
Figure\,\ref{fig:all_candidates_mdwarfs}, we show the folded light
curve of the 3 MEB candidates. The parameters associated with the main
stellar companion of 19a1-02980 \& 19a1-10878 are listed in
Table\,\ref{tab:vosa_mdwarf}. For the candidate 19c4-06354, we report
a primary stellar companion with J-band = 17.97, in a short period
system of $P$\,$\sim$\,0.76 days and low $T_{eff}$ of 3500\,K. Due to
the faint magnitude of the primary stellar companion, this MEB
candidate was found only in the DI light curves. A more extensive and
meticulous search for MEB systems in the DI light curves (which is out
of the scope of this work) could potentially reveal many more
detections in the future.\\

\section{Conclusions}
\label{sec:conclusion}

We carried out a quantitative comparison between the photometric
precision of two different sets of light curves from the 19h field,
which represents the most complete field of the WTS. The light curves
were obtained using two different photometric techniques, AP and
DI. The $sysrem$ algorithm was used to remove systematic effects in
both data sets and corrected the light curves by scaling the error
bars. The WTS AP light curves reach a slightly better photometric
precision (by $\sim$1\,mmag) than the DI light curves for objects
brighter than J\,$\approx$\,15.5\,mag. On the other hand, the DI light
curves show a significant improvement of $\sim$2-20\,mmag for sources
with magnitudes larger than J\,=\,16\,mag.

A modified version of the box-fitting algorithm was employed to search
for transiting planets in the survey. Our algorithm uses the standard
BLS to searches for the best trial period and subsequently makes a
trapezoid re-fitting to the folded light curve, providing a new
estimation of the transit depth. A $\chi^2$ comparison shows that the
new trapezoid fit provides better results than the traditional
box-fitting. The algorithm also calculates a new parameter based on
the geometry of the new trapezoid fit, the $V$-shape parameter. This
parameter has proven to be very efficient in the identification and
removal of eclipsing binaries from the candidate sample.\\

In order to select our candidates, we proposed a set of selection
criteria, 6 of them are based on the experience of previous works.
Additionally, 2 new criteria were incorporated, which take advantage
of the results obtained with our transit detection algorithm, such as
the $V$-shape parameter. The set of parameters of our selection
criteria was optimized using Monte Carlo simulations by injecting
transit signals to both the AP and DI light curves. The light curves
were split in two different sets, one for F-G-K-stars, and a second
for M-dwarfs. The optimization of the criteria was performed in both
sets separately. The selection criteria have shown the capability of
detecting 200 candidates in the DI and AP light curves from a original
sample of $\sim$\,475\,000 F-G-K-stars, while 196 candidates were
detected in both sets of light curves from the M-dwarfs sample. We
carried out a visual examination on the detections and identified
18 relevant transit planet candidates.\\

In order to discriminate planetary from binary candidates, a detailed
analysis of the 18 candidates was conducted, which provides physical
parameters of the candidates and their host stars. The analysis
includes a characterization of the parent star and a transit fit of
the light curve using a realistic model proposed by
\citet{2002ApJ...580L.171M}. Furthermore, we performed a secondary
eclipse fit to the phase folded light curve using the double period to
detect potential differences in the $\chi^2_{dof}$ and/or the depths
of the primary and secondary eclipses that could be an indication of
an eclipsing binary system. In our analysis, only one object is
classified as a planet candidate, which is proposed for photometric
follow-up. The remaining 17 candidates have large best fitting radii
and are therefore classified as binary candidates.\\

No planet candidates orbiting an M-dwarf was found, therefore, the
null detection presented in \citet{2013MNRAS.tmp.1446K} was
confirmed. A detailed sensitivity analysis allowed us to derive an
upper limit on the occurrence rate of giant planets around M-dwarfs
with periods below 10\,days. Increasing the number of target stars by
going one magnitude deeper, we were able to set a 95\,\% confidence
upper limit of 1.1\,\%, which is significantly lower than any limit
published so far. Another applications of the WTS DI light curves were
reported. We presented the detection of five new ultra-short period
eclipsing binaries with periods below 0.23\,days and J\,$>$\,17
mag. In addition, three detached M-dwarf eclipsing binary candidates
were reported; two of them were found in both the AP and DI light
curves, while the third and faintest candidate was only detected in
the DI light curves sample. These results show that the DI light
curves are able to reproduce and improve results reported in the
literature. In conclusion, the WTS DI light curves are useful for many
purposes, such as detection of transit planet candidates and rare
eclipsing binary systems, especially when pushing the limits to
fainter magnitudes.

\clearpage

\clearpage

\clearpage

\clearpage
\newpage
\begin{acknowledgements}
  We acknowledge the support of RoPACS network during this research, a
  Marie Curie Initial Training Network funded by the European
  Commissions Seventh Framework Programme.  This publication makes use
  of VOSA, developed under the Spanish Virtual Observatory project
  supported from the Spanish MICINN through grant AyA2008-02156. This
  research has been also funded by the Spanish grants
  AYA2012-38897-C02-01, PRICIT-S2009/ESP-1496 and
  AyA2011-30147-C03-03. This work has made use of the NASA/IPAC
  Extragalactic Database (NED) which is operated by the Jet Propulsion
  Laboratory, California Institute of Technology, under contract with
  the National Aeronautics and Space Administration. Furthermore, we
  have made use of NASA's Astrophysics Data System, as well as, the
  SIMBAD database operated at CDS, Strasbourg, France.\\
\end{acknowledgements}
\bibliographystyle{aa_chicho.bst}
\bibliography{literature}

\begin{thebibliography}{60}
\expandafter\ifx\csname natexlab\endcsname\relax\def\natexlab#1{#1}\fi

\bibitem[{{Adelman-McCarthy} \& {et al.}(2009)}]{2009yCat.2294....0A}
{Adelman-McCarthy}, J.~K. \& {et al.} 2009, VizieR Online Data Catalog, 2294, 0

\bibitem[{{Aigrain} {et~al.}(2008){Aigrain}, {Barge}, {Deleuil}, {Fressin},
  {Moutou}, {Queloz}, {Auvergne}, \& {Baglin}}]{2008ASPC..384..270A}
{Aigrain}, S., {Barge}, P., {Deleuil}, M., {et~al.} 2008, in Astronomical
  Society of the Pacific Conference Series, Vol. 384, 14th Cambridge Workshop
  on Cool Stars, Stellar Systems, and the Sun, ed. G.~{van Belle}, 270

\bibitem[{{Alard} \& {Lupton}(1998)}]{1998ApJ...503..325A}
{Alard}, C. \& {Lupton}, R.~H. 1998, \apj, 503, 325

\bibitem[{{Baraffe} {et~al.}(1998){Baraffe}, {Chabrier}, {Allard}, \&
  {Hauschildt}}]{1998A&A...337..403B}
{Baraffe}, I., {Chabrier}, G., {Allard}, F., \& {Hauschildt}, P.~H. 1998, \aap,
  337, 403

\bibitem[{{Barge} {et~al.}(2008){Barge}, {Baglin}, {Auvergne}, \& {CoRoT
  Team}}]{2008IAUS..249....3B}
{Barge}, P., {Baglin}, A., {Auvergne}, M., \& {CoRoT Team}. 2008, in IAU
  Symposium, Vol. 249, IAU Symposium, ed. Y.-S. {Sun}, S.~{Ferraz-Mello}, \&
  J.-L. {Zhou}, 3--16

\bibitem[{{Bayo} {et~al.}(2008){Bayo}, {Rodrigo}, {Barrado Y Navascu{\'e}s},
  {Solano}, {Guti{\'e}rrez}, {Morales-Calder{\'o}n}, \&
  {Allard}}]{2008A&A...492..277B}
{Bayo}, A., {Rodrigo}, C., {Barrado Y Navascu{\'e}s}, D., {et~al.} 2008, \aap,
  492, 277

\bibitem[{{Berta} {et~al.}(2012){Berta}, {Irwin}, {Charbonneau}, {Burke}, \&
  {Falco}}]{2012AJ....144..145B}
{Berta}, Z.~K., {Irwin}, J., {Charbonneau}, D., {Burke}, C.~J., \& {Falco},
  E.~E. 2012, \aj, 144, 145

\bibitem[{{Birkby} {et~al.}(2012){Birkby}, {Nefs}, {Hodgkin}, {Kov{\'a}cs},
  {Sip{\H o}cz}, {Pinfield}, {Snellen}, {Mislis}, {Murgas}, {Lodieu}, {de
  Mooij}, {Goulding}, {Cruz}, {Stoev}, {Cappetta}, {Palle}, {Barrado},
  {Saglia}, {Martin}, \& {Pavlenko}}]{2012MNRAS.426.1507B}
{Birkby}, J., {Nefs}, B., {Hodgkin}, S., {et~al.} 2012, \mnras, 426, 1507

\bibitem[{{Birkby} {et~al.}(2013{\natexlab{a}}){Birkby}, {Cappetta}, {Cruz},
  {Koppenhoefer}, {Ivanyuk}, {Mustill}, {Hodgkin}, {Pinfield}, {Sip{\H o}cz},
  {Kov{\'a}cs}, {Saglia}, {Pavlenko}, \& {RoPACS
  Collaboration}}]{2013EPJWC..4701004B}
{Birkby}, J.~L., {Cappetta}, M., {Cruz}, P., {et~al.} 2013{\natexlab{a}}, in
  European Physical Journal Web of Conferences, Vol.~47, European Physical
  Journal Web of Conferences, 1004

\bibitem[{{Birkby} {et~al.}(2013{\natexlab{b}}){Birkby}, {Cappetta}, {Cruz},
  {Koppenhoefer}, {Ivanyuk}, \& {Mustill}}]{2013MNRAS..111..111}
{Birkby}, J.~L., {Cappetta}, M., {Cruz}, P., {et~al.} 2013{\natexlab{b}},
  \mnras, submitted

\bibitem[{{Borucki} {et~al.}(2010){Borucki}, {Koch}, {Basri}, {Batalha},
  {Brown}, {Caldwell}, {Caldwell}, {Christensen-Dalsgaard}, {Cochran},
  {DeVore}, {Dunham}, {Dupree}, {Gautier}, {Geary}, {Gilliland}, {Gould},
  {Howell}, {Jenkins}, {Kondo}, {Latham}, {Marcy}, {Meibom}, {Kjeldsen},
  {Lissauer}, {Monet}, {Morrison}, {Sasselov}, {Tarter}, {Boss}, {Brownlee},
  {Owen}, {Buzasi}, {Charbonneau}, {Doyle}, {Fortney}, {Ford}, {Holman},
  {Seager}, {Steffen}, {Welsh}, {Rowe}, {Anderson}, {Buchhave}, {Ciardi},
  {Walkowicz}, {Sherry}, {Horch}, {Isaacson}, {Everett}, {Fischer}, {Torres},
  {Johnson}, {Endl}, {MacQueen}, {Bryson}, {Dotson}, {Haas}, {Kolodziejczak},
  {Van Cleve}, {Chandrasekaran}, {Twicken}, {Quintana}, {Clarke}, {Allen},
  {Li}, {Wu}, {Tenenbaum}, {Verner}, {Bruhweiler}, {Barnes}, \&
  {Prsa}}]{2010Sci...327..977B}
{Borucki}, W.~J., {Koch}, D., {Basri}, G., {et~al.} 2010, Science, 327, 977

\bibitem[{{Burke} {et~al.}(2006){Burke}, {Gaudi}, {DePoy}, \&
  {Pogge}}]{2006AJ....132..210B}
{Burke}, C.~J., {Gaudi}, B.~S., {DePoy}, D.~L., \& {Pogge}, R.~W. 2006, \aj,
  132, 210

\bibitem[{{Cappetta} {et~al.}(2012){Cappetta}, {Saglia}, {Birkby},
  {Koppenhoefer}, {Pinfield}, {Hodgkin}, {Cruz}, {Kov{\'a}cs}, {Sip{\H o}cz},
  {Barrado}, {Nefs}, {Pavlenko}, {Fossati}, {del Burgo}, {Mart{\'{\i}}n},
  {Snellen}, {Barnes}, {Bayo}, {Campbell}, {Catalan}, {G{\'a}lvez-Ortiz},
  {Goulding}, {Haswell}, {Ivanyuk}, {Jones}, {Kuznetsov}, {Lodieu}, {Marocco},
  {Mislis}, {Murgas}, {Napiwotzki}, {Palle}, {Pollacco}, {Sarro Baro},
  {Solano}, {Steele}, {Stoev}, {Tata}, \& {Zendejas}}]{2012MNRAS.427.1877C}
{Cappetta}, M., {Saglia}, R.~P., {Birkby}, J.~L., {et~al.} 2012, \mnras, 427,
  1877

\bibitem[{{Castelli} {et~al.}(1997){Castelli}, {Gratton}, \&
  {Kurucz}}]{1997A&A...318..841C}
{Castelli}, F., {Gratton}, R.~G., \& {Kurucz}, R.~L. 1997, \aap, 318, 841

\bibitem[{{Charbonneau} {et~al.}(2000){Charbonneau}, {Brown}, {Latham}, \&
  {Mayor}}]{2000ApJ...529L..45C}
{Charbonneau}, D., {Brown}, T.~M., {Latham}, D.~W., \& {Mayor}, M. 2000, \apjl,
  529, L45

\bibitem[{{Claret} \& {Bloemen}(2011)}]{2011A&A...529A..75C}
{Claret}, A. \& {Bloemen}, S. 2011, \aap, 529, A75

\bibitem[{{Defa{\"y}} {et~al.}(2001){Defa{\"y}}, {Deleuil}, \&
  {Barge}}]{2001A&A...365..330D}
{Defa{\"y}}, C., {Deleuil}, M., \& {Barge}, P. 2001, \aap, 365, 330

\bibitem[{{Derekas} {et~al.}(2007){Derekas}, {Kiss}, \&
  {Bedding}}]{2007ApJ...663..249D}
{Derekas}, A., {Kiss}, L.~L., \& {Bedding}, T.~R. 2007, \apj, 663, 249

\bibitem[{{Devor}(2005)}]{2005ApJ...628..411D}
{Devor}, J. 2005, \apj, 628, 411

\bibitem[{{Dimitrov} \& {Kjurkchieva}(2010)}]{2010MNRAS.406.2559D}
{Dimitrov}, D.~P. \& {Kjurkchieva}, D.~P. 2010, \mnras, 406, 2559

\bibitem[{{Dotter} {et~al.}(2008){Dotter}, {Chaboyer}, {Jevremovi{\'c}},
  {Kostov}, {Baron}, \& {Ferguson}}]{2008ApJS..178...89D}
{Dotter}, A., {Chaboyer}, B., {Jevremovi{\'c}}, D., {et~al.} 2008, \apjs, 178,
  89

\bibitem[{{Giacobbe} {et~al.}(2012){Giacobbe}, {Damasso}, {Sozzetti}, {Toso},
  {Perdoncin}, {Calcidese}, {Bernagozzi}, {Bertolini}, {Lattanzi}, \&
  {Smart}}]{2012MNRAS.424.3101G}
{Giacobbe}, P., {Damasso}, M., {Sozzetti}, A., {et~al.} 2012, \mnras, 424, 3101

\bibitem[{{Gould} {et~al.}(2006){Gould}, {Dorsher}, {Gaudi}, \&
  {Udalski}}]{2006AcA....56....1G}
{Gould}, A., {Dorsher}, S., {Gaudi}, B.~S., \& {Udalski}, A. 2006, \actaa, 56,
  1

\bibitem[{{Hartman} {et~al.}(2009){Hartman}, {Gaudi}, {Holman}, {McLeod},
  {Stanek}, {Barranco}, {Pinsonneault}, {Meibom}, \&
  {Kalirai}}]{2009ApJ...695..336H}
{Hartman}, J.~D., {Gaudi}, B.~S., {Holman}, M.~J., {et~al.} 2009, \apj, 695,
  336

\bibitem[{{Henry} {et~al.}(2000){Henry}, {Marcy}, {Butler}, \&
  {Vogt}}]{2000ApJ...529L..41H}
{Henry}, G.~W., {Marcy}, G.~W., {Butler}, R.~P., \& {Vogt}, S.~S. 2000, \apjl,
  529, L41

\bibitem[{{Henry} {et~al.}(1997){Henry}, {Ianna}, {Kirkpatrick}, \&
  {Jahreiss}}]{1997AJ....114..388H}
{Henry}, T.~J., {Ianna}, P.~A., {Kirkpatrick}, J.~D., \& {Jahreiss}, H. 1997,
  \aj, 114, 388

\bibitem[{{Hodgkin} {et~al.}(2009){Hodgkin}, {Irwin}, {Hewett}, \&
  {Warren}}]{2009MNRAS.394..675H}
{Hodgkin}, S.~T., {Irwin}, M.~J., {Hewett}, P.~C., \& {Warren}, S.~J. 2009,
  \mnras, 394, 675

\bibitem[{{Howard} {et~al.}(2012){Howard}, {Marcy}, {Bryson}, {Jenkins},
  {Rowe}, {Batalha}, {Borucki}, {Koch}, {Dunham}, {Gautier}, {Van Cleve},
  {Cochran}, {Latham}, {Lissauer}, {Torres}, {Brown}, {Gilliland}, {Buchhave},
  {Caldwell}, {Christensen-Dalsgaard}, {Ciardi}, {Fressin}, {Haas}, {Howell},
  {Kjeldsen}, {Seager}, {Rogers}, {Sasselov}, {Steffen}, {Basri},
  {Charbonneau}, {Christiansen}, {Clarke}, {Dupree}, {Fabrycky}, {Fischer},
  {Ford}, {Fortney}, {Tarter}, {Girouard}, {Holman}, {Johnson}, {Klaus},
  {Machalek}, {Moorhead}, {Morehead}, {Ragozzine}, {Tenenbaum}, {Twicken},
  {Quinn}, {Isaacson}, {Shporer}, {Lucas}, {Walkowicz}, {Welsh}, {Boss},
  {Devore}, {Gould}, {Smith}, {Morris}, {Prsa}, {Morton}, {Still}, {Thompson},
  {Mullally}, {Endl}, \& {MacQueen}}]{2012ApJS..201...15H}
{Howard}, A.~W., {Marcy}, G.~W., {Bryson}, S.~T., {et~al.} 2012, \apjs, 201, 15

\bibitem[{{Irwin} {et~al.}(2009){Irwin}, {Charbonneau}, {Nutzman}, \&
  {Falco}}]{2009AIPC.1094..445I}
{Irwin}, J., {Charbonneau}, D., {Nutzman}, P., \& {Falco}, E. 2009, in American
  Institute of Physics Conference Series, Vol. 1094, 15th Cambridge Workshop on
  Cool Stars, Stellar Systems, and the Sun, ed. E.~{Stempels}, 445--448

\bibitem[{{Irwin} {et~al.}(2007){Irwin}, {Irwin}, {Aigrain}, {Hodgkin}, {Hebb},
  \& {Moraux}}]{2007MNRAS.375.1449I}
{Irwin}, J., {Irwin}, M., {Aigrain}, S., {et~al.} 2007, \mnras, 375, 1449

\bibitem[{{Irwin} \& {Lewis}(2001)}]{2001NewAR..45..105I}
{Irwin}, M. \& {Lewis}, J. 2001, \nar, 45, 105

\bibitem[{{Irwin}(1985)}]{1985MNRAS.214..575I}
{Irwin}, M.~J. 1985, \mnras, 214, 575

\bibitem[{{Ivezi{\'c}} {et~al.}(2005){Ivezi{\'c}}, {Vivas}, {Lupton}, \&
  {Zinn}}]{2005AJ....129.1096I}
{Ivezi{\'c}}, {\v Z}., {Vivas}, A.~K., {Lupton}, R.~H., \& {Zinn}, R. 2005,
  \aj, 129, 1096

\bibitem[{{Jehin} {et~al.}(2011){Jehin}, {Gillon}, {Queloz}, {Magain},
  {Manfroid}, {Chantry}, {Lendl}, {Hutsem{\'e}kers}, \&
  {Udry}}]{2011Msngr.145....2J}
{Jehin}, E., {Gillon}, M., {Queloz}, D., {et~al.} 2011, The Messenger, 145, 2

\bibitem[{{Jenkins} {et~al.}(2010){Jenkins}, {Chandrasekaran}, {McCauliff},
  {Caldwell}, {Tenenbaum}, {Li}, {Klaus}, {Cote}, \&
  {Middour}}]{2010SPIE.7740E..10J}
{Jenkins}, J.~M., {Chandrasekaran}, H., {McCauliff}, S.~D., {et~al.} 2010, in
  Society of Photo-Optical Instrumentation Engineers (SPIE) Conference Series,
  Vol. 7740, Society of Photo-Optical Instrumentation Engineers (SPIE)
  Conference Series

\bibitem[{{Kaltenegger} \& {Traub}(2009)}]{2009ApJ...698..519K}
{Kaltenegger}, L. \& {Traub}, W.~A. 2009, \apj, 698, 519

\bibitem[{{Kasting} {et~al.}(1993){Kasting}, {Whitmire}, \&
  {Reynolds}}]{1993Icar..101..108K}
{Kasting}, J.~F., {Whitmire}, D.~P., \& {Reynolds}, R.~T. 1993, \icarus, 101,
  108

\bibitem[{{Kleinmann} {et~al.}(1994){Kleinmann}, {Lysaght}, {Pughe},
  {Schneider}, {Skrutskie}, {Weinberg}, {Price}, {Matthews}, {Soifer}, \&
  {Huchra}}]{1994Ap&SS.217...11K}
{Kleinmann}, S.~G., {Lysaght}, M.~G., {Pughe}, W.~L., {et~al.} 1994, \apss,
  217, 11

\bibitem[{{Koppenhoefer} {et~al.}(2009){Koppenhoefer}, {Afonso}, {Saglia}, \&
  {Henning}}]{2009A&A...494..707K}
{Koppenhoefer}, J., {Afonso}, C., {Saglia}, R.~P., \& {Henning}, T. 2009, \aap,
  494, 707

\bibitem[{{Kov{\'a}cs} {et~al.}(2013){Kov{\'a}cs}, {Hodgkin}, {Sip{\H o}cz},
  {Pinfield}, {Barrado}, {Birkby}, {Cappetta}, {Cruz}, {Koppenhoefer},
  {Mart{\'{\i}}n}, {Murgas}, {Nefs}, {Saglia}, \&
  {Zendejas}}]{2013MNRAS.tmp.1446K}
{Kov{\'a}cs}, G., {Hodgkin}, S., {Sip{\H o}cz}, B., {et~al.} 2013, \mnras

\bibitem[{{Kov{\'a}cs} {et~al.}(2002){Kov{\'a}cs}, {Zucker}, \&
  {Mazeh}}]{2002A&A...391..369K}
{Kov{\'a}cs}, G., {Zucker}, S., \& {Mazeh}, T. 2002, \aap, 391, 369

\bibitem[{{Law} {et~al.}(2012){Law}, {Kraus}, {Street}, {Fulton},
  {Hillenbrand}, {Shporer}, {Lister}, {Baranec}, {Bloom}, {Bui}, {Burse},
  {Cenko}, {Das}, {Davis}, {Dekany}, {Filippenko}, {Kasliwal}, {Kulkarni},
  {Nugent}, {Ofek}, {Poznanski}, {Quimby}, {Ramaprakash}, {Riddle},
  {Silverman}, {Sivanandam}, \& {Tendulkar}}]{2012ApJ...757..133L}
{Law}, N.~M., {Kraus}, A.~L., {Street}, R., {et~al.} 2012, \apj, 757, 133

\bibitem[{{L{\'o}pez-Morales} \& {Ribas}(2005)}]{2005ApJ...631.1120L}
{L{\'o}pez-Morales}, M. \& {Ribas}, I. 2005, \apj, 631, 1120

\bibitem[{{Maceroni} \& {Montalb{\'a}n}(2004)}]{2004A&A...426..577M}
{Maceroni}, C. \& {Montalb{\'a}n}, J. 2004, \aap, 426, 577

\bibitem[{{Mandel} \& {Agol}(2002)}]{2002ApJ...580L.171M}
{Mandel}, K. \& {Agol}, E. 2002, \apjl, 580, L171

\bibitem[{{Mayor} \& {Queloz}(1995)}]{1995Natur.378..355M}
{Mayor}, M. \& {Queloz}, D. 1995, \nat, 378, 355

\bibitem[{{Mazeh} {et~al.}(2009){Mazeh}, {Guterman}, {Aigrain}, {Zucker},
  {Grinberg}, {Alapini}, {Alonso}, {Auvergne}, {Barbieri}, {Barge},
  {Bord{\'e}}, {Bouchy}, {Deeg}, {de La Reza}, {Deleuil}, {Dvorak}, {Erikson},
  {Fridlund}, {Gondoin}, {Jorda}, {Lammer}, {L{\'e}ger}, {Llebaria}, {Magain},
  {Moutou}, {Ollivier}, {P{\"a}tzold}, {Pont}, {Queloz}, {Rauer}, {Rouan},
  {Sabo}, {Schneider}, \& {Wuchterl}}]{2009A&A...506..431M}
{Mazeh}, T., {Guterman}, P., {Aigrain}, S., {et~al.} 2009, \aap, 506, 431

\bibitem[{{Meeus}(1982)}]{1982QB51.3.E43M43..}
{Meeus}, J. 1982, {Astronomical formulae for calculators}, 43

\bibitem[{{Montalto} {et~al.}(2007){Montalto}, {Piotto}, {Desidera}, {de
  Marchi}, {Bruntt}, {Stetson}, {Arellano Ferro}, {Momany}, {Gratton},
  {Poretti}, {Aparicio}, {Barbieri}, {Claudi}, {Grundahl}, \&
  {Rosenberg}}]{2007A&A...470.1137M}
{Montalto}, M., {Piotto}, G., {Desidera}, S., {et~al.} 2007, \aap, 470, 1137

\bibitem[{{Nefs} {et~al.}(2012){Nefs}, {Birkby}, {Snellen}, {Hodgkin},
  {Pinfield}, {Sip{\H o}cz}, {Kov\'{a}cs}, {Mislis}, {Saglia}, {Koppenhoefer},
  {Cruz}, {Barrado}, {Martin}, {Goulding}, {Stoev}, {Zendejas}, {del Burgo},
  {Cappetta}, \& {Pavlenko}}]{2012MNRAS.425..950N}
{Nefs}, S.~V., {Birkby}, J.~L., {Snellen}, I.~A.~G., {et~al.} 2012, \mnras,
  425, 950

\bibitem[{{Norton} {et~al.}(2011){Norton}, {Payne}, {Evans}, {West},
  {Wheatley}, {Anderson}, {Barros}, {Butters}, {Collier Cameron}, {Christian},
  {Enoch}, {Faedi}, {Haswell}, {Hellier}, {Holmes}, {Horne}, {Kane}, {Lister},
  {Maxted}, {Parley}, {Pollacco}, {Simpson}, {Skillen}, {Smalley},
  {Southworth}, \& {Street}}]{2011A&A...528A..90N}
{Norton}, A.~J., {Payne}, S.~G., {Evans}, T., {et~al.} 2011, \aap, 528, A90

\bibitem[{{Pietrukowicz} {et~al.}(2010){Pietrukowicz}, {Minniti},
  {D{\'{\i}}az}, {Fern{\'a}ndez}, {Zoccali}, {Gieren}, {Pietrzy{\'n}ski},
  {Ru{\'{\i}}z}, {Udalski}, {Szeifert}, \& {Hempel}}]{2010A&A...509A...4P}
{Pietrukowicz}, P., {Minniti}, D., {D{\'{\i}}az}, R.~F., {et~al.} 2010, \aap,
  509, A4

\bibitem[{{Pont} {et~al.}(2006){Pont}, {Zucker}, \&
  {Queloz}}]{2006MNRAS.373..231P}
{Pont}, F., {Zucker}, S., \& {Queloz}, D. 2006, \mnras, 373, 231

\bibitem[{{Robin} {et~al.}(2003){Robin}, {Reyl{\'e}}, {Derri{\`e}re}, \&
  {Picaud}}]{2003A&A...409..523R}
{Robin}, A.~C., {Reyl{\'e}}, C., {Derri{\`e}re}, S., \& {Picaud}, S. 2003,
  \aap, 409, 523

\bibitem[{{Rucinski}(1992)}]{1992AJ....103..960R}
{Rucinski}, S.~M. 1992, \aj, 103, 960

\bibitem[{{Scalo} {et~al.}(2007){Scalo}, {Kaltenegger}, {Segura}, {Fridlund},
  {Ribas}, {Kulikov}, {Grenfell}, {Rauer}, {Odert}, {Leitzinger}, {Selsis},
  {Khodachenko}, {Eiroa}, {Kasting}, \& {Lammer}}]{2007AsBio...7...85S}
{Scalo}, J., {Kaltenegger}, L., {Segura}, A.~G., {et~al.} 2007, Astrobiology,
  7, 85

\bibitem[{{Snellen} {et~al.}(2007){Snellen}, {van der Burg}, {de Hoon}, \&
  {Vuijsje}}]{2007A&A...476.1357S}
{Snellen}, I.~A.~G., {van der Burg}, R.~F.~J., {de Hoon}, M.~D.~J., \&
  {Vuijsje}, F.~N. 2007, \aap, 476, 1357

\bibitem[{{Tamuz} {et~al.}(2005){Tamuz}, {Mazeh}, \&
  {Zucker}}]{2005MNRAS.356.1466T}
{Tamuz}, O., {Mazeh}, T., \& {Zucker}, S. 2005, \mnras, 356, 1466

\bibitem[{{Tarter} {et~al.}(2007){Tarter}, {Backus}, {Mancinelli}, {Aurnou},
  {Backman}, {Basri}, {Boss}, {Clarke}, {Deming}, {Doyle}, {Feigelson},
  {Freund}, {Grinspoon}, {Haberle}, {Hauck}, {Heath}, {Henry}, {Hollingsworth},
  {Joshi}, {Kilston}, {Liu}, {Meikle}, {Reid}, {Rothschild}, {Scalo}, {Segura},
  {Tang}, {Tiedje}, {Turnbull}, {Walkowicz}, {Weber}, \&
  {Young}}]{2007AsBio...7...30T}
{Tarter}, J.~C., {Backus}, P.~R., {Mancinelli}, R.~L., {et~al.} 2007,
  Astrobiology, 7, 30

\bibitem[{{Tomaney} \& {Crotts}(1996)}]{1996AJ....112.2872T}
{Tomaney}, A.~B. \& {Crotts}, A.~P.~S. 1996, \aj, 112, 2872

\end{thebibliography}

\end{document}